\pdfoutput=1
\documentclass[cernpreprint, atlasdraft=false, texlive=2023, UKenglish, texmf, orcidlogo]{atlasdoc}
 
\usepackage[subcaption]{atlaspackage}
\usepackage[backref=false]{atlasbiblatex}
 
\usepackage{atlasphysics}
\usepackage{atlasbsm}
 
\graphicspath{{logos/}{figures/}}
 
\usepackage{ANA-SUSY-2019-02-PAPER-defs}
\usepackage{multirow}
\usepackage{graphicx}
\usepackage{epstopdf}
\usepackage{rotating}
\usepackage{footnote}
\makesavenoteenv{tabular}
\makesavenoteenv{table}
\newenvironment{DIFnomarkup}{}{}

\AtlasTitle{Search for direct pair production of sleptons and charginos decaying to two leptons and neutralinos with mass splittings near the $W$-boson mass in $\boldmath{\sqrt{s}=13\,}$TeV $pp$ collisions with the ATLAS detector}
 
\AtlasJournalRef{JHEP 06 (2023) 031}
\AtlasDOI{10.1007/JHEP06(2023)031}

\AtlasAbstract{
A search for the electroweak production of pairs of charged sleptons or charginos decaying into two-lepton final states with missing transverse momentum is presented.
Two simplified models of $R$-parity-conserving supersymmetry are considered: direct pair-production of sleptons ($\tilde{\ell}\tilde{\ell}$), with each decaying into a charged lepton and a $\tilde{\chi}_1^0$ neutralino, and direct pair-production of the lightest charginos $(\tilde{\chi}_1^\pm\tilde{\chi}_1^\mp)$, with each decaying into a $W$-boson and a $\tilde{\chi}_1^0$.
The lightest neutralino ($\tilde{\chi}_1^0$) is assumed to be the lightest supersymmetric particle (LSP).
The analyses target the experimentally challenging mass regions where $m(\tilde{\ell})-m(\tilde{\chi}_1^0)$ and $m(\tilde{\chi}_1^\pm)-m(\tilde{\chi}_1^0)$ are close to the $W$-boson mass (`moderately compressed' regions). The search uses 139~fb$^{-1}$ of $\sqrt{s}=13$~TeV proton--proton collisions recorded by the ATLAS detector at the Large Hadron Collider.
No significant excesses over the expected background are observed.
Exclusion limits on the simplified models under study are reported in the ($\tilde{\ell},\tilde{\chi}_1^0$) and ($\tilde{\chi}_1^\pm,\tilde{\chi}_1^0$) mass planes at 95\% confidence level (CL). Sleptons with masses up to 150~GeV are excluded at 95\% CL for the case of a mass-splitting between sleptons and the LSP of 50~GeV. Chargino masses up to 140~GeV are excluded at 95\% CL for the case of a mass-splitting between the chargino and the LSP down to about 100~GeV.
}
 
\author{The ATLAS Collaboration}

\AtlasRefCode{SUSY-2019-02}

\AtlasJournal{JHEP}
 
\PreprintIdNumber{CERN-EP-2022-132}

\hypersetup{pdftitle={ATLAS document},pdfauthor={The ATLAS Collaboration}}
\begin{document}
 
\maketitle

\section{Introduction}
 
Weak-scale supersymmetry (SUSY)~\cite{Golfand:1971iw,Volkov:1973ix,Wess:1974tw,Wess:1974jb,Ferrara:1974pu,Salam:1974ig} is a theoretical extension of the Standard Model (SM) which can solve the fine-tuning problem through the addition of a new fermionic/bosonic supersymmetric partner to each boson/fermion in the SM.
In SUSY models with  $R$-parity conservation~\cite{Farrar:1978xj}, SUSY particles must be produced in pairs and the lightest supersymmetric particle (LSP) is stable and weakly interacting, thus being a candidate for dark matter ~\cite{Goldberg:1983nd,Ellis:1983ew}.
 
The SUSY particle production cross-sections at the Large Hadron Collider (LHC) are highly dependent on their masses.
Squarks and gluinos are strongly produced and have significantly larger production \mbox{cross-sections}
than non-coloured SUSY particles of equal masses, such as the sleptons (superpartners of the SM leptons) and the electroweakinos (the higgsino, wino and bino superpartners of the SM Higgs boson and electroweak gauge bosons).
The electroweakinos mix to form chargino ($\tilde{\chi}_i^\pm, i=1,2$) and neutralino ($\tilde{\chi}_j^{0}, j=1,2,3,4$) mass eigenstates, where the states are ordered by increasing values of their mass.
 
The presence of light scalar partners of the SM leptons just above the mass of the LSP is still not excluded in an interesting, but experimentally challenging, area in the $m(\tilde\ell)$--$m(\tilde{\chi}_1^0)$ plane.
In particular, electroweak-scale SUSY with light smuons (superpartners of the SM muons) and a light LSP can explain the $(g-2)_\mu$ anomaly~\cite{Bennett:2006,Abi:2021} through additional loop corrections.
The $(g-2)_\mu$ anomaly favours the mass region in the $m(\tilde{\mu})$--$m(\tilde{\chi}_1^0)$ plane that is `moderately compressed' or `compressed' (i.e.\ the region where the mass splitting $m(\tilde{\mu})-m(\tilde{\chi}_1^0)$ is close to, or smaller than, the $W$-boson mass) for small $\tan\beta$ values\footnote{In the Minimal Supersymmetric Standard Model (MSSM) $\tan\beta$ is defined as the ratio of the vacuum expectation values of the two complex Higgs doublets.}~\cite{Heinemeyer:2006}.
For small $\tan\beta$, results from LEP and previous LHC searches~\cite{ATL-PHYS-PUB-2022-013} excluded portions of this mass region, but an important part of it is still not ruled out. This search extends the sensitivity to light smuons beyond that reached by LEP and previous LHC searches.
 
The analyses presented in this paper target the direct production of slepton pairs decaying into the LSP via the emission of a charged lepton, and the direct production of $\tilde{\chi}_1^+\tilde{\chi}_1^-$, where each chargino decays to the LSP via the emission of a $W$-boson, which decays leptonically. A signature with two opposite-charge leptons (electrons and/or muons), \MET (defined as the magnitude of the missing transverse momentum $\mathbf p^\mathrm{miss}_\mathrm{T}$)
and low hadronic activity is considered. A moderately compressed mass spectrum is targeted.
 
A previous search \cite{SUSY-2018-32} by ATLAS considered the same models and signature.
The search exploited the full ATLAS Run~2 data set, but it was optimized to target the phase-space with a large mass
difference between the chargino or slepton and the LSP. The event selection was based on the two-lepton invariant mass, the \MET, the \MET\ significance~\cite{ATLAS-CONF-2018-038}, a veto against $b$-tagged jets (i.e.\ those originating from $b$-quarks), and a requirement of less than two light jets. Finally, a shape-fitting technique was applied, exploiting several bins of the \mttwo\ distribution.\footnote{The \mttwo\ variable is defined in Section~\ref{sec:kine_var}}
 
The new searches in this paper complement previous ones, extending the sensitivity in the experimentally challenging phase-space regions with mass splittings $m(\tilde{\ell})-m(\tilde{\chi}_1^0)$ and $m(\tilde{\chi}_1^\pm)-m(\tilde{\chi}_1^0)$ close to the $W$-boson mass, which result in topologies very similar to those of SM processes.
The areas of parameter space excluded by these results extend beyond those
excluded by previous searches by ATLAS~\cite{SUSY-2013-11,SUSY-2016-24,SUSY-2018-32} and CMS~\cite{CMS-SUS-13-006,CMS-SUS-14-002,CMS-SUS-16-039,CMS-SUS-17-004,CMS-SUS-17-009,CMS-SUS-17-010} in the same channels.
 
The gain in sensitivity is due to a dedicated analysis strategy for each of the two signal scenarios considered.
Since the slepton signal presents only a same-flavour lepton-pair signature, the SM background is estimated with a data-driven technique using events with two leptons of different flavour and opposite-sign electric charge.
In the chargino search, the signal results in both same-flavour and different-flavour lepton pairs and
the topology of the signal is close to the SM $WW$ process. In this case,
a machine-learning technique is used, based on boosted decision trees specifically trained on signal samples with $m(\tilde{\chi}_1^\pm)-m(\tilde{\chi}_1^0)$ around the $W$-boson mass.
 
This paper is structured as follows. Section~\ref{sec:physics_scenarios} and Section~\ref{sec:detector} describe the signal scenarios considered in these searches and the ATLAS detector, respectively.
The data and simulated Monte Carlo (MC) samples used in the analyses, along with the trigger selections, are detailed in Section~\ref{sec:DataMC}. Section~\ref{sec:EvRec} describes the physics object definitions.
The search strategies and the SM background estimations are discussed in Sections~\ref{sec:search} and~\ref{sec:backgrounds}, respectively.
The experimental and theoretical systematic uncertainties considered in the two searches are documented in Section~\ref{sec:Syst}. Finally, the results and their statistical interpretations are presented in Section~\ref{sec:result}, followed by the conclusion in Section~\ref{sec:conclusion}.
 
\section{SUSY scenarios}
\label{sec:physics_scenarios}
The design of the analyses and the interpretation of the results are based on simplified models~\cite{Alwall:2008ve,Alwall:2008ag,Alves:2011wf}, where the masses of relevant sparticles (in this case the $\tilde{\ell}$, $\chinoonepm$ and $\ninoone$) are the only free parameters and all the other sparticles are assumed to be heavy and decoupled.
 
\begin{figure}[htb!]
\centering
\begin{subfigure}{0.40\linewidth}
\includegraphics[width=\linewidth]{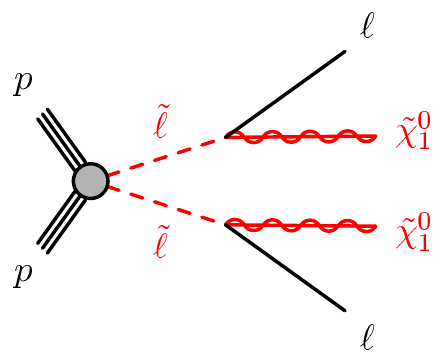}
\caption{}
\end{subfigure}
\begin{subfigure}{0.40\linewidth}
\includegraphics[width=\linewidth]{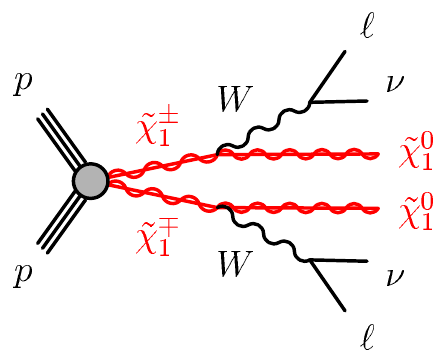}
\caption{}
\end{subfigure}
\caption{Diagrams of the supersymmetric simplified models considered, with two charged leptons plus weakly interacting particles in the final state: (a)~slepton pair production and
(b)~$\tilde{\chi}_1^+\tilde{\chi}_1^-$ production with $W$-boson-mediated decays.
Only $\tilde{e}$ and $\tilde{\mu}$ are included in the slepton model.
In the final state, $\ell$ stands for an electron or muon, which can be produced directly or, in the case of (b), via a leptonically decaying $\tau$-lepton along with additional neutrinos.
}
\label{fig:c1c1viawwfeynman}
\end{figure}

In models with direct $\tilde\ell\tilde\ell$ production (Figure~\ref{fig:c1c1viawwfeynman}(a)), each slepton decays into a charged lepton and a bino-like $\tilde{\chi}_1^0$ with a 100\% branching ratio. Only $\tilde{e}$ and $\tilde{\mu}$ are considered in these models, and different assumptions about the masses of the superpartners of the left-handed and right-handed charged leptons, $\tilde{e}_\mathrm{L}$, $\tilde{e}_\mathrm{R}$, $\tilde{\mu}_\mathrm{L}$ and $\tilde{\mu}_\mathrm{R}$,
are considered.
Lepton flavour is conserved in all models.
 
The $\tilde{\chi}_1^\pm$ is assumed to be wino-like and decay into a bino-like $\tilde{\chi}_1^0$ via emission of a $W$-boson, which may decay into an electron or muon plus neutrino(s) either directly or through the emission of a leptonically decaying $\tau$-lepton (Figure~\ref{fig:c1c1viawwfeynman}(b)).

\FloatBarrier
 
\section{ATLAS detector}
\label{sec:detector}

The ATLAS detector~\cite{PERF-2007-01} at the LHC is a multipurpose particle detector with a forward--backward symmetric cylindrical geometry and an almost complete coverage in solid angle around the collision point.\footnote{{
ATLAS uses a right-handed coordinate system with its origin at the nominal interaction point (IP) in the centre of the detector and the $z$-axis along the beam pipe.
The $x$-axis points from the IP to the centre of the LHC ring, and the $y$-axis points upwards.
Cylindrical coordinates $(r,\phi)$ are used in the transverse plane, $\phi$ being the azimuthal angle around the $z$-axis.
The pseudorapidity is defined in terms of the polar angle $\theta$ as $\eta = -\ln \tan(\theta/2)$. Rapidity is defined as $y=(1/2)\ln[(E+p_z)/(E-p_z)]$, where $E$ and $p_z$ denote the energy and the component of the particle momentum along the beam direction, respectively.
}}
It consists of an inner tracking detector surrounded by a thin superconducting solenoid providing a 2\,T axial magnetic field, electromagnetic and hadron calorimeters, and a muon spectrometer.
 
The inner-detector (ID) system covers the pseudorapidity range $|\eta| < 2.5$. It consists of a high-granularity silicon pixel tracker, a silicon microstrip tracker, and a straw-tube transition-radiation tracker, which enables radially extended track reconstruction up to $|\eta| = 2.0$ and provides electron identification information.
The insertable $B$-layer (IBL)~\cite{ATLAS-TDR-19,PIX-2018-001}, installed before Run~2, typically records the innermost hit on a track.
 
The calorimeter system covers the pseudorapidity range $|\eta| < 4.9$.
Lead/liquid-argon (LAr) sampling calorimeters provide electromagnetic (EM) energy measurements with high granularity within the region $|\eta|< 3.2$. An iron/scintillator-tile hadron calorimeter covers the central pseudorapidity range $|\eta| < 1.7$.
The endcap and forward regions are instrumented with LAr calorimeters for EM and hadronic energy measurements.
 
The muon spectrometer (MS) surrounds the calorimeters and incorporates three large air-core toroidal superconducting magnets with eight coils each, providing a field integral ranging between 2.0 and 6.0~Tm across most of the detector.
It comprises a system of precision tracking chambers measuring the deflection of muons in the magnetic field and fast detectors for triggering.
The precision chamber system covers the region $|\eta| < 2.7$,
while the muon trigger system covers the range $|\eta| < 2.4$.
 
A two-level trigger system is used to select events. The first-level (L1) trigger is implemented in hardware and accepts events at a maximum rate of 100~kHz using a subset of detector information. It is followed by a software-based high-level trigger (HLT), which reduces the accepted event rate to 1~kHz on average depending on the data-taking conditions, selecting interesting events with algorithms accessing the full detector information~\cite{TRIG-2016-01,TRIG-2019-04}. An extensive software suite~\cite{ATL-SOFT-PUB-2021-001} is used in the reconstruction and analysis of real and simulated data, in detector operations, and in the trigger and data acquisition systems of the experiment.

\section{Data and simulated event samples}
\label{sec:DataMC}
 
The data set used in these analyses was collected by the ATLAS detector in $pp$ collisions provided by the LHC during Run~2 from 2015 to 2018.
The beams collided at a centre-of-mass energy of $\sqrt{s} = 13$~\TeV\ and with a minimum separation of 25~ns between consecutive proton bunch  crossings.
The average number $\langle \mu \rangle$ of additional $pp$ interactions per bunch crossing (pile-up) ranged from 14 in 2015 to about 38 in 2017--2018.
After data-quality requirements~\cite{DAPR-2018-01}, applied to ensure that all parts of the detector were operational during data-taking, the data sample amounts to a total integrated luminosity of 139~fb$^{-1}$.
The uncertainty in the combined 2015--2018 integrated luminosity is 1.7\%~\cite{ATLAS-CONF-2019-021}, obtained using the LUCID-2 detector~\cite{LUCID2} for the primary luminosity measurements.
 
Candidate events were selected by various single-electron or single-muon triggers~\cite{TRIG-2018-01,TRIG-2018-05}.
The transverse momentum (\pT) thresholds of the single-lepton triggers were raised as the luminosity increased during Run~2.  They were in the range 20--120~\GeV\ for data collected in 2015, 24--300~\GeV\ for data collected in 2016, and 26--300~\GeV\ for data collected in 2017 and 2018. Higher \pT requirements were applied to the lepton involved in the trigger decision during the offline lepton selection to ensure that trigger efficiencies are constant in the relevant phase-space.
 
MC generators were used to simulate samples of collision events, which model the expected kinematics of the investigated signal and SM background processes.
For background processes, the detector response was simulated using the full modelling of the ATLAS detector~\cite{SOFT-2010-01} in \GEANT~\cite{Agostinelli:2002hh}, while for the signal samples a faster version of the simulation was used, which relies on a parameterization for the response of the calorimeters and on \GEANT for the other components of the detector~\cite{SOFT-2010-01}.
The effect of pile-up was modelled by overlaying the hard-scatter events with simulated inelastic $pp$ events generated with \PYTHIA[8.186]~\cite{Sjostrand:2008} and \EVTGEN~\cite{Lange:2001uf}, using the \NNPDF[2.3lo] set of
parton distribution functions (PDF)~\cite{Ball:2012cx} and the A3 set of tuned parameters~\cite{ATL-PHYS-PUB-2016-017}. The MC samples were reweighted so that the distribution of the average number of interactions per bunch crossing reproduces the observed distribution in the data.
All simulated events were processed with the same trigger, reconstruction and identification algorithms as the data. Correction factors were applied to the simulated event samples to account for differences between data and simulation in the jet and lepton reconstruction efficiencies, energy scales, and energy resolutions, and in the lepton trigger efficiency~\cite{EGAM-2018-01,PERF-2015-10} and jet $b$-tagging efficiency~\cite{FTAG-2018-01}.
 
Table~\ref{tab:mc_backgr} gives a detailed summary of all SM background samples used in the analyses. It lists the generators, the PDF sets, the sets of underlying-event and hadronisation parameter values (tune) for the parton shower, and the order of the cross-section computation in $\alphas$. Further information about ATLAS simulations of \ttbar, single-top ($Wt$), multiboson and boson-plus-jet processes can be found in the relevant public notes~\cite{ATL-PHYS-PUB-2018-009,ATL-PHYS-PUB-2016-004,ATL-PHYS-PUB-2017-005,ATL-PHYS-PUB-2017-006}.

\begin{DIFnomarkup}
\begin{table}[!h]
\tiny
\centering
\caption{\label{tab:mc_backgr}Simulated background event samples with the corresponding matrix element and parton shower (PS) generators, cross-section order in $\alphas$ used to normalize the event yield, underlying-event tune and the generator PDF sets used. For diboson, triboson and $t\bar{t}{+}V$ samples, $V=W,Z$. Diboson samples also include Higgs boson contributions. `Default' refers to the \SHERPA generator's default tune.}
\begin{tabular}{lp{0.18\linewidth}p{0.143\linewidth}p{0.123\linewidth}p{0.059\linewidth}p{0.104\linewidth}p{0.11\linewidth}}
\toprule
 
Physics process          			& Generator 	& Parton shower 	&  Normalization 		& Tune 			& PDF (generator) & PDF (PS)\\
\midrule
 
$t\bar{t}$ 				& \POWHEGBOX[v2]~\cite{Frixione:2007nw,Nason:2004rx,Frixione:2007vw,Alioli:2010xd}       		& \PYTHIA[8.230]~\cite{Sjostrand:2014zea}  		& NNLO+NNLL~\cite{Beneke:2011mq,Cacciari:2011hy,Baernreuther:2012ws,Czakon:2012zr,Czakon:2012pz,Czakon:2013goa,Czakon:2011xx} 		& A14~\cite{ATL-PHYS-PUB-2014-021} & \NNPDF[3.0nlo]~\cite{Ball:2014uwa} & \NNPDF[2.3lo]~\cite{Ball:2012cx}\\
Single-top ($Wt$) 				& \POWHEGBOX[v2]~\cite{Re:2010bp,Nason:2004rx,Frixione:2007vw,Alioli:2010xd} 	& \PYTHIA[8.230]	&NLO+NNLL~\cite{Kidonakis:2010ux,Kidonakis:2013zqa}		& A14 			& \NNPDF[3.0nlo]
& \NNPDF[2.3lo]
\\
Diboson $VV$			&  \SHERPA[2.2.1],\,2.2.2~\cite{Bothmann:2019yzt}	& 	\SHERPA[2.2.1],\,2.2.2~\cite{Gleisberg:2008fv,Schumann:2007mg}
& NLO~\cite{Hoeche:2011fd,Hoeche:2012yf,Catani:2001cc,Hoeche:2009rj}   		&  Default~\cite{ATL-PHYS-PUB-2017-005}
&\NNPDF[3.0nnlo]~\cite{Ball:2014uwa}   & \NNPDF[3.0nnlo] \\
Triboson $VVV$				& \SHERPA[2.2.2]			&  \SHERPA[2.2.2]		& NLO			& Default 	& \NNPDF[3.0nnlo]
& \NNPDF[3.0nnlo]
\\
$t\bar{t}{+}V$ 			& \MGNLO[2.3.3]~\cite{Alwall:2014hca}			& \PYTHIA[8.210]~\cite{Sjostrand:2014zea}						& NLO~\cite{Alwall:2014hca} 			& A14 & \NNPDF[3.0nlo]
& \NNPDF[2.3lo]
\\
$t\bar{t}{+}H$ 				& \POWHEGBOX[v2]~\cite{Frixione:2007nw,Nason:2004rx,Frixione:2007vw,Alioli:2010xd,Hartanto:2015uka}      		& \PYTHIA[8.230] 		& NLO	& A14 & \NNPDF[3.0nlo] & \NNPDF[2.3lo]\\
$t\bar{t}{+}WW$				& \MGNLO[2.2.2]			& \PYTHIA[8.186]~\cite{Sjostrand:2008} 		& NLO~\cite{Alwall:2014hca}			& A14			& \NNPDF[2.3lo]
&  \NNPDF[2.3lo]
\\  
$t\bar{t}{+}WZ$				& \MGNLO[2.3.3]			& \PYTHIA[8.212]~\cite{Sjostrand:2014zea}
& NLO~\cite{Alwall:2014hca}			& A14			& \NNPDF[2.3lo]
&  \NNPDF[2.3lo]
\\  
$tZ,t\bar{t}t\bar{t},t\bar{t}t$		& \MGNLO[2.3.3]			& \PYTHIA[8.230]
& NLO~\cite{Alwall:2014hca}      		& A14			& \NNPDF[3.0nlo]
& \NNPDF[2.3lo]
\\
$Z/\gamma^*(\rightarrow \ell\ell )$+jets 	& \SHERPA[2.2.1]~\cite{Bothmann:2019yzt}			& \SHERPA[2.2.1]~\cite{Schumann:2007mg}		& NNLO~\cite{Anastasiou:2003ds}			& Default	& \NNPDF[3.0nnlo] & \NNPDF[3.0nnlo] \\

$H\rightarrow\tau\tau$           &\POWHEGBOX[v2]~\cite{Nason:2004rx,Frixione:2007vw,Alioli:2010,Hamilton:2013fea,Hamilton:2015nsa}                    & \PYTHIA[8.2]~\cite{Sjostrand:2014zea}                  &NNLO~\cite{Hamilton:2012np,Campbell:2012am,Hamilton:2012rf,Catani:2007vq}              &AZNLO~\cite{STDM-2012-23}      & \NNPDF[3.0nlo]   & \PDFforLHC[15nnlo]~\cite{Butterworth:2015oua} \\

\bottomrule
 
\end{tabular}
 
\end{table}
 
The SUSY signal samples were generated from leading-order (LO) matrix elements with up to two extra partons, using \MADGRAPH[v2.6.1]~\cite{madgraph} for direct $\tilde\ell\tilde\ell$ production and \MADGRAPH[v2.6.2]~\cite{madgraph} for $\tilde{\chi}_1^+\tilde{\chi}_1^-\rightarrow W^+\tilde{\chi}_1^0W^-\tilde{\chi}_1^0$,
interfaced with \PYTHIA[8.244]~\cite{Sjostrand:2014zea} and \PYTHIA[8.212]~\cite{Sjostrand:2014zea} respectively, with the A14 set of tuned parameters~\cite{ATL-PHYS-PUB-2014-021}, for the modelling of the SUSY decay chain, parton showers, hadronization and underlying event. In order to include spin correlation effects in off-shell $W$-boson decays, \MADSPIN~\cite{Artoisenet:2012st} was used in the event generation for mass-splittings between the chargino and LSP smaller than 100~\GeV.
Parton luminosities were provided by the \NNPDF[2.3lo] PDF set~\cite{Ball:2012cx}. Jet--parton matching followed the CKKW-L prescription~\cite{Lonnblad:2012ix}, with a matching scale set to one quarter of the pair-produced superpartner mass for the slepton model and to 15~\GeV\ for the chargino model.
Signal cross-sections were calculated to next-to-leading order (NLO) in $\alphas$, with resummation of soft gluon emission at next-to-leading-logarithm (NLL) accuracy~\cite{Debove:2010kf,Fuks:2012qx,Fuks:2013vua,Fiaschi:2018hgm,Bozzi:2007qr_B,Fuks:2013lya,Fiaschi:2018xdm}.
The nominal cross-sections and their uncertainties were taken from an envelope of cross-section predictions using different PDF sets and factorization and renormalization scales, as described in Ref.~\cite{Borschensky:2014cia}.
The cross-section for $\tilde{\chi}_1^+\tilde{\chi}_1^- $ production, for $m(\tilde\chi_1^\pm) = 150$~\GeV,
is $2.61\pm 0.14$~pb, while the cross-section for ${\tilde\ell\tilde\ell}$ production, for $m(\tilde\ell) = 150$~\GeV,
is $63.3\pm 3.3$~fb for each generation of left-handed sleptons and $23.3\pm 1.4$~fb for each generation of right-handed sleptons.

\section{Object reconstruction}
\label{sec:EvRec}
 
Candidate events are required to have at least one $pp$ interaction vertex with a minimum of two associated tracks, each with $p_\mathrm{T}>500$~\MeV. In events with multiple vertices, the primary vertex is defined as the one with the highest scalar sum of the squared transverse momenta of associated tracks.
 
The leptons selected for the analyses are classified as baseline or signal leptons using an increasingly
stringent set of quality and kinematic selection criteria. The signal leptons are a subset
of the baseline leptons. Baseline objects are used in the calculation of missing transverse momentum, to resolve ambiguities between the analysis objects in the event and in the fake/non-prompt (FNP) lepton background estimation described in Section~\ref{sec:backgrounds}.
Signal leptons are used for the final event selection.
 
Baseline electron candidates are reconstructed using three-dimensional clusters of energy deposits in the electromagnetic calorimeter
that are matched to an ID track. They are required to pass a \textit{Loose} likelihood-based
identification requirement~\cite{EGAM-2018-01} with an additional condition on the number of hits in the pixel detector's innermost layer, and to have $p_\mathrm{T}>9$~\GeV\ and  $|\eta|<2.47$. The tracks associated with baseline electron candidates are required to be within $|z_0 \sin\theta| = 0.5$~mm of the primary vertex, where $z_0$ is the longitudinal impact parameter relative to the reconstructed primary vertex.
Signal electrons are required to satisfy a \textit{Tight} identification requirement~\cite{EGAM-2018-01} and the track associated with the signal electron is required to have $\vert d_0\vert/\sigma(d_0) < 5$, where $d_0$ is  the transverse impact parameter relative to the primary vertex and $\sigma(d_0)$ is its uncertainty.
 
Baseline muon candidates are reconstructed in the pseudorapidity range $|\eta|<2.6$ by matching MS tracks with ID tracks.
They are required to have $p_\mathrm{T}>\SI{9}{\GeV}$, to be within $|z_0 \sin\theta|$ = 0.5~mm of the primary vertex and to satisfy the \textit{Medium} identification requirements defined in Ref.~\cite{PERF-2015-10}, based on the numbers of hits in the different ID and MS subsystems, and on the significance of the charge-to-momentum ratio $q/p$.
Signal muons are required to have the associated track with $\vert d_0\vert/\sigma(d_0) < 3$.
 
Isolation criteria are applied to signal electrons and muons in order to suppress contributions from photon conversions, semileptonic decays of heavy-flavour hadrons, or hadrons and jets wrongly identified as leptons, collectively referred as fake or non-prompt leptons. The scalar sum of the \pt of tracks inside a variable-size cone around the lepton (excluding its own track), must be less than 15\% of the lepton $p_\mathrm{T}$. The track isolation cone size for electrons (muons), $\Delta R=\sqrt{(\Delta\eta)^2+(\Delta\phi)^2}$, is given by the minimum of $\Delta R = \SI{10}{~\GeV}/\pt$ and $\Delta R = 0.2\,(0.3)$.
In addition, for electrons (muons) the sum of the transverse energy ($E_\mathrm{T}= E/ \cosh(\eta)$ from Ref.~\cite{EGAM-2018-01}) of the calorimeter energy clusters in a cone of $\Delta R = 0.2$ around the lepton (excluding the energy from the lepton itself) must be less than 20\% (30\%) of the lepton $p_\mathrm{T}$.
 
Jets are reconstructed from particle-flow objects~\cite{PERF-2015-09} calibrated at the EM scale. The anti-$k_t$ jet clustering algorithm~\cite{Cacciari:2008} as implemented in the FastJet package~\cite{Fastjet} is used with a radius parameter $R=0.4$. The reconstructed jets are corrected to particle level by the application of a jet energy scale (JES) and resolution (JER) calibrations, derived from 13~\TeV\ data and simulation~\cite{PERF-2016-04}. Only jet candidates with $p_\mathrm{T}>\SI{20}{\GeV}$ and $|\eta|<2.4$ are considered. To reduce the effects of pile-up, for jets with $\pt<\SI{60}{\GeV}$ a significant fraction of the tracks associated with each jet are required to have an origin compatible with the primary vertex, as defined by the jet vertex tagger~\cite{ATLAS-CONF-2014-018}. This requirement reduces the fraction of jets from pile-up to 1\%, with an efficiency for pure hard-scatter jets of about 90\%.
Finally, in order to remove events impacted by detector noise and non-collision backgrounds, specific jet-quality requirements~\cite{ATLAS-CONF-2015-029,PERF-2014-03} are applied to ensure that the efficiency of selecting jets from proton--proton collisions is above 99.5\% (99.9\%) for $p_\mathrm{T}>20$~(100)~\GeV.
 
Jets that are likely to have originated from the hadronization of a bottom quark are flagged as `$b$-jets' if they lie within  $|\eta|<2.4$ and are tagged by the DL1r algorithm~\cite{FTAG-2018-01}, a multivariate discriminant based on various inputs such as track impact parameters and displaced secondary vertices. A selection that provides 85\% efficiency for tagging $b$-jets in simulated $t\bar{t}$ events is used. The corresponding rejection factors for jets originating from $c$-quarks, from $\tau$-leptons, and from light quarks and gluons in the same sample at this working point are 2, 4 and 31, respectively.
 
The identities of reconstructed objects may be ambiguous. To prevent single detector signatures from being identified as multiple objects, an overlap-removal procedure is applied to baseline leptons and jets in several consecutive steps:
\begin{itemize}
\item jet candidates within $\Delta R' =\sqrt{\Delta y^2+\Delta\phi^2}$ = 0.2  of an electron candidate, or jets with fewer than three tracks and that lie within $\Delta R' = 0.4$ of a muon candidate, are removed because they mostly originate from calorimeter energy deposits from electron showers or muon bremsstrahlung;
\item electrons and muons within $\Delta R' =$ min$(0.4, 0.04 + 10{~\GeV}/p_\mathrm{T}) $ of the surviving jets are discarded, to reject leptons from the decay of $b$- or $c$-hadrons;
\item if an electron shares an ID track with a muon, the electron is discarded unless the muon is tagged as a minimum-ionizing particle in the calorimeter, in which case the muon is discarded.
\end{itemize}
 
The missing transverse momentum $\mathbf p^\mathrm{miss}_\mathrm{T}$ is defined as the negative vector sum of the transverse momenta of all identified baseline physics objects (electrons, photons, muons and jets), and an additional soft term including all tracks that pass basic quality requirements and are associated with the primary vertex but not with any reconstructed physics object~\cite{PERF-2016-07}. The magnitude of $\mathbf p^\mathrm{miss}_\mathrm{T}$ is denoted by \MET. Additionally, an `object-based \MET\ significance'~\cite{ATLAS-CONF-2018-038}, referred to as \MET\ significance in this paper, helps to discriminate between events where \MET arises from undetected particles in the final state and those where it arises from poorly measured particles, the $\mathbf p_\mathrm{T}$ resolution, or identification inefficiencies. It is defined as
\begin{equation*}
E^\mathrm{miss}_\mathrm{T}\ \mathrm{significance} = \frac{|\mathbf p^\mathrm{miss}_\mathrm{T}|}{\sqrt{\sigma_\text{L}^2(1-\rho_\text{LT}^2)}}
\end{equation*}
where $\sigma_\text{L}$ is the longitudinal component (parallel to the $\mathbf p^\mathrm{miss}_\mathrm{T}$) of the total transverse momentum resolution for all objects in the event and the quantity $\rho_\text{LT}$ is the correlation factor between the parallel and perpendicular
components of the transverse momentum resolution for each object.

\section{Event selection}
\label{sec:search}
 
\subsection{Preselection}
 
The strategy for event preselection uses a common approach for the two analysis models, and is described here. Further selections, specific for each of the two target scenarios, are discussed in Sections~\ref{sec:slept_search} and~\ref{sec:charg_search}.
 
Events are required to have exactly two oppositely charged signal leptons $\ell_1$ and $\ell_2$,  with $\ell_1$ having $p_\mathrm{T} > 27~\GeV$ (leading lepton) and $\ell_2$ having $p_\mathrm{T} > 9~\GeV$ (sub-leading lepton). The invariant mass of the two leptons must satisfy $m_{\ell\ell}>11$~\GeV, in order to remove low-mass resonances.
 
Events are separated into two classes, `same-flavour' (SF) events, i.e.\ $e^{\pm}e^{\mp}$ and $\mu^{\pm}\mu^{\mp}$, and `different-flavour' (DF) events, i.e.\  $e^{\pm}\mu^{\mp}$, since they have different background compositions. SF events are required to have a dilepton invariant mass far from the $Z$ peak,  $|m_{\ell\ell} - 91~\GeV| > 15$~\GeV,  to reduce $VZ$  ($V=W,Z$) and $Z$+jets backgrounds. Events must also have no more than one jet ($n_{\mathrm{jet}}< 2$) and satisfy \MET\ significance~$>3$.
 
\subsection{Kinematic variables}
\label{sec:kine_var}
Final event selections are performed by separating signal from SM background using different kinematic variables. For both SUSY models in Figure~\ref{fig:c1c1viawwfeynman}, the stransverse mass $m_{\mathrm{T2}}$~\cite{Lester:1999tx,Barr:2003rg} and $\cos\theta^{*}_{\ell\ell}$, defined below, are among the most discriminating variables.  The stransverse mass generalizes the transverse mass\footnote{The transverse mass is defined as
$m^2_\textup{T} = m^2_1 + m^2_2 + 2(E_{\mathrm{T,1}}E_{\mathrm{T,2}} + \mathbf p_{\mathrm{T,1}} \cdot \mathbf p_{\mathrm{T,2}})$,
where $m_1$,$m_2$ are the masses and $\mathbf p_{\mathrm{T,1}}$,$\mathbf p_{\mathrm{T,2}}$ are the transverse momenta of the two particles.}
$m_{\mathrm T}$ for symmetric event topologies where two identical particles each decay into a visible and an invisible product. In this case the individual transverse momenta of the invisible particles can no longer be directly approximated by the measured missing transverse momentum, because the information about their individual contributions to the missing transverse momentum is lost. The stransverse mass is defined as
\begin{equation*}
\mttwo
( \mathbf p_{\mathrm{T,1}}, \mathbf p_{\mathrm{T,2}},\mathbf p^\mathrm{miss}_\mathrm{T}) = \min_{\mathbf q_{\mathrm{T,1}} + \mathbf q_{\mathrm{T,2}} =
\mathbf p^\mathrm{miss}_\mathrm{T} }
\left\{ \max[\;
m_{\mathrm{T}}( \mathbf p_{\mathrm{T,1}}, \mathbf q_{\mathrm{T,1}} ),
m_{\mathrm{T}}( \mathbf p_{\mathrm{T,2}}, \mathbf q_{\mathrm{T,2}}
)
\;]
\right\},
\end{equation*}
 
\noindent where $\mathbf p_{\mathrm{T,1}}$ and $\mathbf p_{\mathrm{T,2}}$ are the transverse-momentum vectors of the two leptons, and $\mathbf
q_{\mathrm{T,1}}$ and $\mathbf q_{\mathrm{T,2}}$ are vectors with
$\mathbf p^\mathrm{miss}_\mathrm{T}= \mathbf q_{\mathrm{T,1}} + \mathbf q_{\mathrm{T,2}}$.
The minimization is performed over all the possible decompositions of $\mathbf p^\mathrm{miss}_\mathrm{T}$. The masses of the invisible particles are free parameters and are set to $100$~\GeV\ (defining $m_{\mathrm{T2}}^{100}$) in the slepton search and to $0$~\GeV\ (defining $m_{\mathrm{T2}}^{0}$) in the chargino search, since these choices improve the sensitivity to several signal models in the mass range targeted by the respective analyses.
 
The variables $m_{\mathrm{T2}}^{100}$ and $m_{\mathrm{T2}}^{0}$, typically having different kinematic endpoints for SUSY processes and for SM processes such as $t\bar{t}$ or $WW$, provide powerful discrimination between background events and some of the signals considered herein.
 
The angular variable $\cos\theta^{*}$, where $\theta^{*}$ is the polar angle between the incoming quark in one of the protons and the produced sparticle, is sensitive to the sparticle spin, and the cross-section behaves differently for scalar sparticles like sleptons, spin-1 sparticles or spin-1/2 sparticles.
Since  $\theta^{*}$ is not directly measurable, $\cos\theta^{*}_{\ell\ell} = \tanh(\Delta\eta_{\ell\ell}/2)$ is defined in terms of the pseudorapidity difference between the two leptons. In the slepton model, it is sensitive to the slepton production angle. The leptons `inherit' some knowledge of the rapidity of their slepton parents, and the two variables $\cos\theta^{*}$ and $\cos\theta^{*}_{\ell\ell}$ are well correlated with each other~\cite{Barr:2005dz}.
Other variables providing powerful discrimination between signal events and backgrounds such as \ttbar or $VV$ are the azimuthal angular separations between the two leptons, \dphill, between $\mathbf p^\mathrm{miss}_\mathrm{T}$ and the leading lepton, \dphiMETlone, and between $\mathbf p^\mathrm{miss}_\mathrm{T}$ and the sub-leading lepton, \dphiMETltwo. The distributions of these variables are affected by the presence of jets in the event.
For example, in slepton pair production in the absence of jets, the sleptons are expected to be produced back to back in the azimuthal plane, and the leptons coming from their decays to be well separated there. The most energetic \ninoone and the sub-leading lepton are expected to come from the same slepton, so the $\mathbf p^\mathrm{miss}_\mathrm{T}$ vector is expected to be well separated from the $\mathbf p_\mathrm{T}^{\ell_1}$ direction.
 
Another exploited variable is the magnitude of $\mathbf\pbll$ ($\pbll$), the vector sum of the $\mathbf p_\mathrm{T}$ of the two leptons and $\mathbf p^\mathrm{miss}_\mathrm{T}$. It can be interpreted as the
magnitude of the vector sum of all the transverse hadronic activity in the event. In both of the analysed SUSY scenarios, in absence of jets,
$\pbll$ is expected to have low values due to the $p_\mathrm{T}$ balance of the system. The azimuthal separation between $\mathbf p^\mathrm{miss}_\mathrm{T}$ and $\mathbf\pbll$ is denoted by \dphib.

\subsection{Signal regions}
Dedicated signal-enriched regions (SRs) are defined for each signal scenario, optimized individually for benchmark signal models by maximizing the discovery significance. The selection requirements for the signal regions are explained in the following for the slepton and chargino scenarios. In both cases, they target signal models with a moderate mass difference between slepton/chargino and neutralino, up to $\sim150$~\GeV.
 
\subsubsection{Slepton model}
\label{sec:slept_search}
 
The event selection which targets the slepton model requires a SF opposite-charge (SFOS) lepton pair, \MET coming from the LSPs, and low hadronic activity apart from initial-state radiation (ISR) or pile-up.
No dedicated selection for $\tilde{e}_\mathrm{L}$, $\tilde{e}_\mathrm{R}$, $\tilde{\mu}_\mathrm{L}$ or $\tilde{\mu}_\mathrm{R}$ is performed.
After the preselection, only events with $n_{\textrm{}b\textrm{-tagged jets}}=0$, i.e.\ the number of jets identified as $b$-jets by the DL1r  algorithm, are retained, in order to reduce the \ttbar and single-top backgrounds.
Events are then further classified by the multiplicity of non-$b$-tagged jets (0J,1J).
 
Following the classification of the events, a dedicated optimization for each of the two categories is performed. A relevant difference between them is related to the requirements on $\pbll$ and $\dphiMETlone$, which are no longer useful for the 1J event category since the presence of the jet implies that the \MET and the leptons' $p_\mathrm{T}$ are not balanced anymore.
Then, two sets of SRs are defined: a set of exclusive SRs, `binned' in $m_{\mathrm{T2}}^{100}$, and a set of `inclusive' SRs, to be used for model-dependent and model-independent results, respectively.
The binning in $m_{\mathrm{T2}}^{100}$ is chosen to maximize the search sensitivity and retain a sufficient number of events in each bin, and the `inclusive' SRs are defined for various lower bounds on $m_{\mathrm{T2}}^{100}$ to enhance sensitivity to new physics with various mass scales.
The definitions of these regions are shown in Table~\ref{tab:SRslepdefs}. Each SR is identified by the number of non-$b$-tagged jets (0J,1J)
and the range of the $m_{\mathrm{T2}}^{100}$ interval.

\begin{table}[!htb]
\caption{\label{tab:SRslepdefs}
The definitions of the binned and inclusive signal regions for the slepton model. Relevant kinematic variables are defined in the text.
The `0J' and `1J' labels refer to the multiplicity of non-$b$-tagged jets.
}
\vspace{-0.25cm}
\begin{center}
\small
\begin{tabular*}{0.7\textwidth}{@{\extracolsep{\fill}}l|ll}
\noalign{\smallskip}\hline\hline\noalign{\smallskip}
\multicolumn{1}{l}{} & &   \\[\dimexpr-\normalbaselineskip-\arrayrulewidth]
Signal region (SR)                       & SR-0J & SR-1J   \\
\noalign{\smallskip}\hline\noalign{\smallskip}
$n_{b \text{-tagged jets}}$               & \multicolumn{2}{c}{= 0} \\
\MET\ significance             & \multicolumn{2}{c}{$>$7} \\
 
\noalign{\smallskip}\hline\noalign{\smallskip}
$n_{\text{non-} b \text{-tagged jets}}$                       & = 0 & = 1    \\
 
\noalign{\smallskip}\hline\noalign{\smallskip}
${p_\mathrm{T}}^{\ell_1}$ [\GeV]			       &  $>$ 140 & 		$>$ 100 \\
${p_\mathrm{T}}^{\ell_2}$ [\GeV]			       &  $>$ 20   & 		$>$ 50 \\
$m_{\ell\ell}$ [\GeV]                  & $>$ 11     &             $>$ 60 \\
$\pbll$	[\GeV] 			       & $<$ 5  &             	-         \\
$|\cos\theta^{*}_{\ell\ell}|$		       & $<$ 0.2  &             $<$ 0.1 \\
$\dphill$					       & $>$ 2.2  &             $>$ 2.8 \\
$\dphiMETlone$			       & $>$ 2.2  &             - \\
\noalign{\smallskip}\hline\hline\noalign{\smallskip}
Binned SRs & &  \\
\noalign{\smallskip}\hline\noalign{\smallskip}
\multirow{8}{*}{$m_{\mathrm{T2}}^{100}$ [\GeV]} & \multicolumn{2}{c}{   \footnotesize $\in$[100,105) }  \\
& \multicolumn{2}{c}{\footnotesize $\in$[105,110) } \\
& \multicolumn{2}{c}{\footnotesize $\in$[110,115) }  \\
& \multicolumn{2}{c}{\footnotesize $\in$[115,120) }  \\
& \multicolumn{2}{c}{\footnotesize $\in$[120,125) }  \\
& \multicolumn{2}{c}{\footnotesize $\in$[125,130) }  \\
& \multicolumn{2}{c}{\footnotesize $\in$[130,140) }  \\
& \multicolumn{2}{c}{\footnotesize $\in$[140,$\infty$)~~ }  \\
\noalign{\smallskip}\hline\hline\noalign{\smallskip}
Inclusive SRs & &       \\
\noalign{\smallskip}\hline\noalign{\smallskip}
\multirow{4}{*}{$m_{\mathrm{T2}}^{100}$ [\GeV]} & \multicolumn{2}{c}{\footnotesize $\in$[100,$\infty$) }  \\
& \multicolumn{2}{c}{\footnotesize $\in$[110,$\infty$) }  \\
& \multicolumn{2}{c}{\footnotesize $\in$[120,$\infty$) }  \\
& \multicolumn{2}{c}{\footnotesize $\in$[130,$\infty$) }  \\
& \multicolumn{2}{c}{\footnotesize $\in$[140,$\infty$) }  \\
\noalign{\smallskip}\hline\hline\noalign{\smallskip}
\end{tabular*}
\end{center}
\end{table}
 
\end{DIFnomarkup}
 
\subsubsection{Chargino model}
\label{sec:charg_search}
The event selection which targets the chargino model considers both same-flavour and different-flavour opposite-charge lepton pairs in the event. After the preselection, only events with $n_{\textrm{}b\textrm{-tagged jets}}=0$ and $n_{\text{non-} b \text{-tagged jets}}=0$ are retained. The first requirement reduces the \ttbar and single-top backgrounds, and the second one was observed to increase the sensitivity of the analysis.
A machine-learning (ML) technique based on the Gradient Boosted Decision Tree (BDT) is exploited in the search for charginos \cite{Ke:2017LightGBM}.
Events passing the preselection and the requirements on the number of jets are separated into two categories, SF and DF, and for each category the signal and SM background Monte Carlo samples are split into two sets: the training set and test set. The
BDT classifier is trained on the training set, and tested on the statistically independent test set. The test set is used to measure and optimize the classifier's performance depending on the parameters which are defined in the ML procedure, and to derive the final results.
Signal samples with a mass-splitting between the chargino and neutralino of 90 or 100~\GeV\ were found to be the best optimization benchmark across the signal grid. They were summed and a part was used for the training set.
Multiclass classification is performed, i.e.\ the classifier is trained to separate events into four classes: signal, $VV$, top (\ttbar and single-top) and all
other backgrounds ($Z/\gamma$+jets, $VVV$ and other minor backgrounds). For each event, the four scores BDT-signal, BDT-VV, BDT-top and BDT-other, corresponding to the four classes, provide the probability for the event to belong to each class, and sum to one. This technique is found to
be more effective than a simpler binary classification in discriminating signal from background.
The set of variables used in the training was optimized in the analysis through an iterative procedure which started from a larger set of variables, removed variables one-by-one and retrained, keeping only variables whose removal caused a loss in performance.
The reduced, final set of variables consists of $p^{\ell_{1}}_\mathrm{T}$, $p^{\ell_{2}}_\mathrm{T}$, \MET, $m_{\mathrm{T2}}^{0}$, \mll, \dphib, \dphiMETlone, \dphiMETltwo, \cosTstar, and \MET\ significance.

\begin{table}[!htb]
\renewcommand{\arraystretch}{1.1}
\caption{\label{tab:SRchargdefs}
The definitions of the binned and inclusive signal regions for the chargino model. Relevant variables are defined in the text. The signal regions are separated for DF and SF, except for the first inclusive SR (subsequently indicated with SR$^{\text{-DF BDT-signal}\in(0.81,1]}_{\text{-SF BDT-signal}\in(0.77,1]}$), which contains DF events with BDT-signal $\in$(0.81,1] and SF events with BDT-signal $\in$(0.77,1].
}
\vspace{-0.25cm}
\begin{center}
\small
\begin{tabular*}{0.7\textwidth}{@{\extracolsep{\fill}}l|ll}
\noalign{\smallskip}\hline\hline\noalign{\smallskip}
\multicolumn{1}{l}{} & &   \\[\dimexpr-\normalbaselineskip-\arrayrulewidth]
Signal region (SR)                       & SR-DF & SR-SF  \\
\noalign{\smallskip}\hline\noalign{\smallskip}
$n_{b \text{-tagged jets}}$                           & \multicolumn{2}{c}{= 0} \\
$n_{\text{non-} b \text{-tagged jets}}$    	& \multicolumn{2}{c}{= 0} \\
\MET\ significance             				& \multicolumn{2}{c}{$>$8} \\
$m_{\mathrm{T2}}^{0}$ [\GeV]            				& \multicolumn{2}{c}{~~$>$50} \\
BDT-other							& 			& $<$ 0.01 \\
 
\noalign{\smallskip}\hline\hline\noalign{\smallskip}
Binned SRs &   \multicolumn{2}{c}{} \\
\noalign{\smallskip}\hline\noalign{\smallskip}
 
\multirow{16}{*}{BDT-signal} & {\footnotesize $\in$(0.81,0.8125] } & {\footnotesize $\in$(0.77,0.775] }\\
& {\footnotesize $\in$(0.8125,0.815] }  & {\footnotesize $\in$(0.775,0.78] }\\
& {\footnotesize $\in$(0.815,0.8175] }  & {\footnotesize $\in$(0.78,0.785] }\\
& {\footnotesize $\in$(0.8175,0.82] }  & {\footnotesize $\in$(0.785,0.79] }\\
& {\footnotesize $\in$(0.82,0.8225] }  & {\footnotesize $\in$(0.79,0.795] }\\
& {\footnotesize $\in$(0.8225,0.825] } & {\footnotesize $\in$(0.795,0.80] } \\
& {\footnotesize $\in$(0.825,0.8275] } & {\footnotesize $\in$(0.80,0.81] } \\
& {\footnotesize $\in$(0.8275,0.83] } & {\footnotesize $\in$(0.81,1] } \\
& {\footnotesize $\in$(0.83,0.8325] }  & \\
& {\footnotesize $\in$(0.8325,0.835] }  & \\
& {\footnotesize $\in$(0.835,0.8375] }  & \\
& {\footnotesize $\in$(0.8375,0.84] }  & \\
& {\footnotesize $\in$(0.84,0.845] }  & \\
& {\footnotesize $\in$(0.845,0.85] }  & \\
& {\footnotesize $\in$(0.85,0.86] }  & \\
& {\footnotesize $\in$(0.86,1] }  & \\
\noalign{\smallskip}\hline\hline\noalign{\smallskip}
Inclusive SRs &   &  \\
\noalign{\smallskip}\hline\noalign{\smallskip}
\multirow{10}{*}{BDT-signal}  & \multicolumn{2}{c}{\footnotesize $\in$(0.81,1] for DF and $\in$(0.77,1] for SF}\\
& {\footnotesize $\in$(0.81,1] }  & 						\\
& {\footnotesize $\in$(0.82,1] }  & 						\\
& {\footnotesize $\in$(0.83,1] }  & 						\\
& {\footnotesize $\in$(0.84,1] }  & 						\\
& {\footnotesize $\in$(0.85,1] }  & 						\\
&   					      & {\footnotesize $\in$(0.77,1] }\\
&   					      & {\footnotesize $\in$(0.78,1] }\\
&   					      & {\footnotesize $\in$(0.79,1] }\\
&   					      & {\footnotesize $\in$(0.80,1] }\\
 
\noalign{\smallskip}\hline\hline\noalign{\smallskip}
\end{tabular*}
\end{center}
\end{table}

Requirements are placed on the BDT scores to define the SRs. Two additional requirements of  \MET\ significance $> 8$ and $m_{\mathrm{T2}}^{0} > 50$~\GeV\ are used in all relevant regions, in order to enhance the sensitivity of the search.
Two sets of signal regions are defined: a set of exclusive SRs, `binned' in BDT-signal to maximize model-dependent search sensitivity, and a set of `inclusive' SRs, to be used for model-independent results,
with a varying lower bound on BDT-signal. The definitions of these regions are shown in Table~\ref{tab:SRchargdefs}.
 
\FloatBarrier

\section{Background estimation}
\label{sec:backgrounds}
 
The SM backgrounds can be classified into irreducible backgrounds, from processes with prompt leptons which can yield events with a final state similar to the signal, and reducible backgrounds, which contain one or more FNP leptons. Among the irreducible backgrounds, for both the slepton and chargino searches the dominant sources are processes with top quarks or dibosons $(VV)$.
 
The slepton search uses a dedicated data-driven technique to estimate some of the dominant backgrounds. This technique is based on the observation that, while the slepton decays produce events with two SFOS leptons in the final state, the decays of background processes such as \ttbar, single-top, $WW$ and $Z(\rightarrow\tau\tau)$+jets produce opposite-sign SF or DF leptons with the same probability (`flavour-symmetric backgrounds', FSB).
The DF channel (populated by the background only) can be used to predict the contribution of FSB to the SF channel (populated by the background and, potentially, by the signal).
 
The chargino search uses a partially data-driven technique to estimate the dominant backgrounds. Dedicated control regions (CRs), enriched in particular backgrounds, are used to normalize MC simulation yields to data.
A simultaneous profile likelihood fit (described in Section~\ref{sec:result}) is used to constrain the MC yields with the observed data. The CRs are designed to be both orthogonal and similar to the SRs, whilst also having little signal contamination; this is achieved by taking the SR definitions and inverting some of the selection criteria. Dedicated validation regions (VRs) are defined to be kinematically close to CRs and SRs, and are used to assess the quality of the background estimation and its extrapolation to the SRs.
 
Subdominant irreducible SM background contributions arising from Drell--Yan, $\ttbar$+boson(s), $tZ$, $t\bar{t}t\bar{t},t\bar{t}t$, Higgs boson, and $VVV$ processes, jointly referred to as `Other backgrounds' (or `Others' in the figures), are estimated from simulation using the samples described in Section~\ref{sec:DataMC}.
 
The reducible background from FNP leptons is estimated from data using the matrix method (MM)~\cite{ATLAS-CONF-2014-058}. This method uses two types of leptons: `signal' leptons, corresponding to leptons passing the final selection used in the analysis, and `baseline' leptons,  which pass a looser selection as defined in Section~\ref{sec:EvRec}.
Probabilities for prompt leptons satisfying the baseline selection to also satisfy the signal selection are measured as a function of lepton \pt and $\eta$ in MC simulation, using control samples enriched in real leptons.
Similar probabilities for FNP leptons to pass the signal selection are measured in data events dominated by leptons from the decays of heavy-flavour hadrons and from photon conversions, and in MC control samples dominated by leptons from light-flavour quark decays. The final probability is then computed by summing the FNP contributions from the different sources, with appropriate weights ($w_i$) which reflect the relative amount from each source, extracted from MC simulations by using event information from the generator's `truth' record.
These probabilities are used in the MM when solving a set of equations relating the numbers of observed baseline and signal leptons to the estimated numbers of real and FNP leptons in the CRs, VRs, and SRs.
To avoid double counting between the simulated samples used for background estimation and the FNP lepton background estimate provided by the MM, all simulated events containing one or more FNP leptons are removed from the background samples.

\subsection{Estimation of the backgrounds in the slepton search} 
 
The number of data events with DF leptons surviving the SR selection ($N_{\mathrm{DF}}$) can be used, after subtracting the FNP lepton contribution, to predict the FSB in the SF channel. Since electrons and muons have different acceptances and trigger, reconstruction, isolation and identification efficiencies, these differences must be taken into account. An efficiency correction method is applied, allowing the expected number of FSB events in the SF channel to be computed as
 
\begin{equation}
\label{eq:SFpredictedAlpha}
\begin{split}
N^{\mathrm{expected}}_{ee} &= 0.5 \times \frac{1}{\kappa} \times \alpha \times N_{\mathrm{DF}} \\
N^{\mathrm{expected}}_{\mu\mu} &= 0.5 \times \kappa \times \alpha \times N_{\mathrm{DF}} \\
N^{\mathrm{expected}}_{\mathrm{SF}} &= 0.5 \times \left(\kappa + \frac{1}{\kappa}\right) \times \alpha \times N_{\mathrm{DF}}
\end{split}
\end{equation}
where it is assumed that efficiencies factorize for the individual leptons and the production rate of the DF events is twice that of the dimuon or dielectron events.
Differences in muon and electron acceptance, identification efficiency, and reconstruction efficiency are taken into account by $\kappa$. Differences in trigger efficiencies are taken into account by $\alpha$.
They are defined as
\begin{equation*}
\begin{split}
\kappa  & = \sqrt{\frac{N_{\mu^{+}\mu^{-}}}{N_{e^{+}e^{-}}}} \\
\alpha & = \frac{\sqrt{\epsilon^{\mathrm{trig}}_{\mu\mu}\epsilon^{\mathrm{trig}}_{ee}}}{\epsilon^{\mathrm{trig}}_{e\mu}}
\end{split}
\end{equation*}
with $N_{\mu^{+}\mu^{-}}$ and $N_{e^{+}e^{-}}$ being the numbers of dimuon and dielectron events respectively, and $\epsilon^{\mathrm{trig}}_{\mu\mu},\epsilon^{\mathrm{trig}}_{ee}$ and $\epsilon^{\mathrm{trig}}_{e\mu}$ the
efficiencies of triggering dimuon, dielectron and electron--muon events with the trigger selection described in Section~\ref{sec:DataMC}.
 
The factor $\kappa$ is extracted from data in a control sample obtained by relaxing the requirements on ${p_\mathrm{T}}^{{\ell}_{1}}$ and the \MET\ significance and inverting the requirement on $|\cos\theta^{*}_{\ell\ell}|$ to make it orthogonal to the SRs. The value of $\kappa$ was observed to depend on the leading lepton's $p_\mathrm{T}$ in different $\eta$ regions of the detector, and it is therefore parameterized as a function of ${p_\mathrm{T}}^{\ell_1}$ only, $\kappa = a+b/p_\mathrm{T}^{\ell_1}$. This parameterization gives a good description of the behaviour of $\kappa$ in both data and MC simulation.
 
The factor $\alpha$ is computed from the global efficiencies of the trigger selection applied in the analysis, evaluated in a control sample of data events passing an independent trigger selection based on  \MET. In the $\eta$ and $p_\mathrm{T}$ ranges where the two leptons satisfy the selection criteria in the SRs, the dependence of $\alpha$ on these kinematic variables was found to be negligible.
 
\begin{figure}[htb!]
\centering
\includegraphics[width=\linewidth]{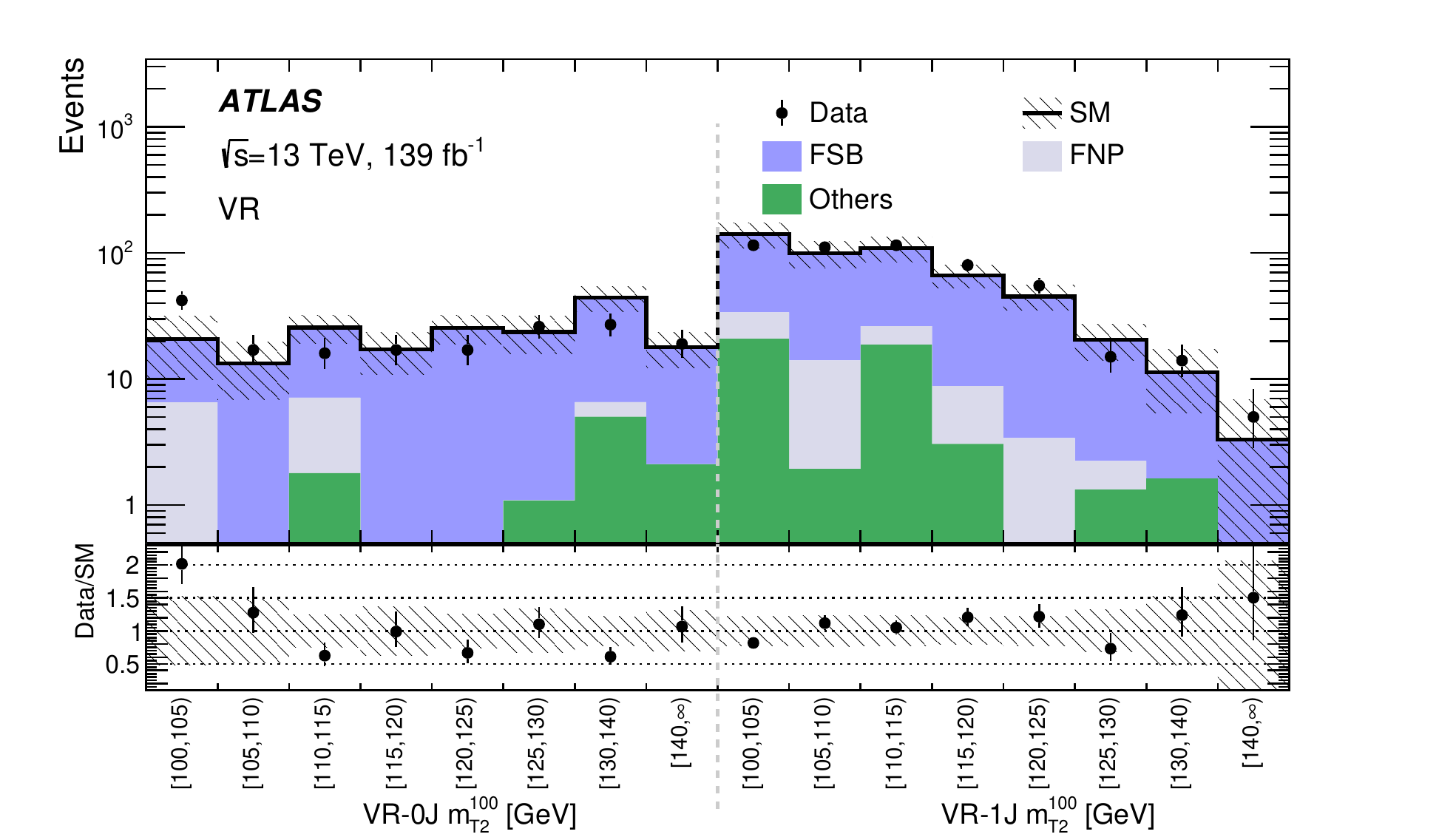}
\caption{Distributions of $m_{\mathrm{T2}}^{100}$ in VR-0J and in VR-1J for data and  the estimated SM backgrounds. The FSB contribution is evaluated with the data-driven efficiency correction method. The FNP lepton background is calculated using the data-driven matrix method. `Others' include the non-dominant background sources, e.g.\ $t \bar{t}$+$V$, Higgs boson and Drell--Yan events. The shaded band represents the total uncertainty, coming from all sources, of the expected SM background. The lower panel shows the ratio of data to the SM background estimate.}
\label{figval_slep}
\end{figure}

In order to validate the efficiency correction method, two validation regions, VR-0J and VR-1J, are defined, with the same selection as in the corresponding SR but inverting the $|\cos\theta^{*}_{\ell\ell}|$ requirement. Although VR-0J and VR-1J are subsets of the control sample used to extract the factor $\kappa$, they use different events, since in these VRs the FSB contribution is evaluated from DF events in data using Eq.~(\ref{eq:SFpredictedAlpha}). Figure~\ref{figval_slep} shows $m_{\mathrm{T2}}^{100}$ in VR-0J and VR-1J, and good agreement
is observed between the data and the total estimated SM background in these distributions and in all other variables relevant for the analysis.

Finally, the FSB yields in SR-0J and SR-1J defined in Table~\ref{tab:SRslepdefs} are estimated using the DF events surviving in data after imposing the selections for each SR and applying the factors $\kappa$ and $\alpha$ on an event-by-event basis. They are reported in Table~\ref{tab:SRexpYields}. \\

\begin{table}[h!]
\caption{Expected flavour-symmetric background yields in SR-0J and SR-1J, estimated from surviving DF data events after imposing the selections for each SR and applying the factors $\kappa$ and $\alpha$ on an event-by-event basis. Yields are separated for $ee$ and $\mu\mu$ events. The uncertainties include both the statistical and systematic contributions.
}
\begin{center}
\begin{tabular*}{0.8\textwidth}{@{\extracolsep{\fill}}lccc}
\noalign{\smallskip}\hline\noalign{\smallskip}
SR & $ee$ events & $\mu\mu$ events & Total \\
\noalign{\smallskip}\hline\noalign{\smallskip}
SR-0J & $ 34.6 \pm 4.9$ & $30.2 \pm 4.4$ & $64.8 \pm 9.3$   \\
SR-1J & $  37.1 \pm 5.0$ & $31.8 \pm 4.5 $ & $ 68.9 \pm 9.4 $   \\
 
\hline
\end{tabular*}
\end{center}
\label{tab:SRexpYields}
\end{table}
 
\FloatBarrier
The irreducible SM non-flavour-symmetric background contribution in SR-0J and SR-1J is estimated directly from simulation using the samples described in Section~\ref{sec:DataMC}.

\subsection{Estimation of the backgrounds in the chargino search} 
 
The general strategy to define CRs and VRs relies on reversing the BDT-signal requirement applied to the SRs or selecting events
with $n_{b \text{-tagged jets}}=1$ for the top CR, in order to ensure orthogonality with the SRs and low signal contamination.
A summary of the regions considered is given in Table~\ref{tab:table_CRVR_charg} and the strategy is described in the following.
 
Two CRs are used, CR-VV to target the diboson backgrounds and CR-top to target the top-quark backgrounds (\ttbar and $Wt$). The selection \MET\ significance $> 8$, $m_{\mathrm{T2}}^{0} > 50$~\GeV\ and $n_{\text{non-} b \text{-tagged jets}} = 0$ applied in these CRs is the same as used in the SRs, in order to ensure that they and the SRs have similar kinematic phase-spaces.
Upper bounds on the BDT-signal score are exploited to ensure orthogonality with the SRs and low signal contamination.
Requirements on the background BDT scores are then applied to ensure that the CR is sufficiently pure in the targeted background.
The $n_{b \text{-tagged jets}}=1$ selection is used in CR-top to ensure a large top-quark background contribution.
A dedicated selection is used for DF and SF events, in order to be consistent with the SR definitions.
The CR selections are summarized in the first part of Table~\ref{tab:table_CRVR_charg}.
The expected number of signal events in the CRs, from the considered models, is at most 5\% of the total background yield.
 
Diboson and top-quark backgrounds are normalized to the data observed in CR-VV and CR-top in a simultaneous likelihood fit,
using a normalization factor for each background ($\mu_\mathrm{VV}$ and $\mu_\mathrm{top}$).
The number of data events observed in each CR, as well as the predicted yield of each SM process, is shown in Table~\ref{tab:CRresults}.
For backgrounds whose normalization is extracted from the likelihood fit, the yield expected from the MC simulation is also reported.
The normalization factors applied to the $VV$ and top-quark backgrounds are found to be $\mu_\mathrm{VV} = 1.38 \pm 0.08$ and $\mu_\mathrm{top}= 1.09 \pm 0.03$, where the errors include all uncertainties described in Section~\ref{sec:Syst}.
The shapes of kinematic distributions are well reproduced by the simulation in each CR, as shown in Figure~\ref{fig:CR}.

\begin{DIFnomarkup}
\begin{table}[!t]
\renewcommand{\arraystretch}{1.2}
\caption{Control region definitions for extracting the normalization factors for the dominant background processes in the chargino search
and validation region definitions used to study the modelling of the SM backgrounds.
The requirements are applied on top of the preselection. `DF' or `SF' refer to control/validation regions with different lepton flavour or same lepton flavour pair combinations, respectively.
}
\begin{center}
\footnotesize
\begin{tabular*}{\textwidth}{@{\extracolsep{\fill}}l|rrrr}
\noalign{\smallskip}\hline\noalign{\smallskip}
Control region (CR) 			&	\multicolumn{2}{c}{CR-VV} 		& \multicolumn{2}{c}{CR-top} \\

\noalign{\smallskip}\hline\noalign{\smallskip}
 
\MET\ significance             				& \multicolumn{4}{c}{$>$ 8} \\
$m_{\mathrm{T2}}^{0}$ [\GeV]            				& \multicolumn{4}{c}{$>$ 50} \\
$n_{\text{non-} b \text{-tagged jets}}$ 		&\multicolumn{4}{c}{$=$ 0}\\
\noalign{\smallskip}\hline\noalign{\smallskip}
Leptons flavour                                                      &	DF & SF	& DF	&SF \\
$n_{b \text{-tagged jets}}$                           & = 0 		& = 0		& = 1		& = 1	\\
BDT-other							& -			& $<$ 0.01 	& -			& $<$ 0.01\\
BDT-signal				&$\in(0.2,0.65]$    &$\in(0.2,0.65]$		&$\in(0.5,0.7]$	&$\in(0.7,0.75]$\\
BDT-VV							&$>$ 0.2		&$>$ 0.2		&-			&-\\
BDT-top							&$<$ 0.1		&$<$ 0.1		&-			&-\\
\noalign{\smallskip}\hline\noalign{\smallskip}
\end{tabular*}
 
\begin{tabular*}{\textwidth}{@{\extracolsep{\fill}}l|rrrrrr}
\noalign{\smallskip}\hline\noalign{\smallskip}
Validation region (VR) 		&VR-VV-DF		&VR-VV-SF		&VR-top-DF		&VR-top-SF   &VR-top0J-DF		&VR-top0J-SF                \\
\noalign{\smallskip}\hline\noalign{\smallskip}
\MET\ significance             				& \multicolumn{6}{c}{$>$ 8} \\
$m_{\mathrm{T2}}^{0}$ [\GeV]            				& \multicolumn{6}{c}{$>$ 50} \\
$n_{\text{non-} b \text{-tagged jets}}$ 	&\multicolumn{6}{c}{$=$ 0}\\
\noalign{\smallskip}\hline\noalign{\smallskip}
 
$n_{ b \text{-tagged jets}}$				& $= 0$		& $= 0$             & $= 1$             &$= 1$ 			&$= 0$ 			&$= 0$ \\
BDT-other							& -			& $<$ 0.01 	& -			& $<$ 0.01		& -				& $<$ 0.01\\
BDT-signal						&$\in(0.65,0.81]$    &$\in(0.65,0.77]$		&$\in(0.7,1]$	&$\in(0.75,1]$ &$\in(0.5,0.81]$	&$\in(0.5,0.77]$\\                                  BDT-VV							&$>$ 0.2		&$>$ 0.2		&-			&-				&$<$ 0.15			&$<$ 0.15\\
BDT-top							&$<$ 0.1		&$<$ 0.1		&-			&-				&-				&-\\
 
\noalign{\smallskip}\hline\noalign{\smallskip}
\end{tabular*}
 
\end{center}
\label{tab:table_CRVR_charg}
\end{table}
\end{DIFnomarkup}
 
\begin{DIFnomarkup}
\begin{table}
\caption{
Observed event yields and predicted background yields from the likelihood fit in the CRs for the chargino search. For backgrounds with a normalization extracted from the likelihood fit, the expected yield from the simulation before the likelihood fit is also shown. The FNP lepton background is calculated using the data-driven matrix method. `Other backgrounds' include the non-dominant background sources, e.g.\ $t\bar{t}$+$V$, Higgs boson and Drell--Yan events. The uncertainties include both the statistical and systematic contributions.
}
\centering
\small
\begin{tabular*}{\textwidth}{@{\extracolsep{\fill}}lcc}
\noalign{\smallskip}\hline\noalign{\smallskip}
Region           & CR-VV            & CR-top              \\
\noalign{\smallskip}\hline\noalign{\smallskip}
Observed events          & $634$              & $4468$                    \\
\noalign{\smallskip}\hline\noalign{\smallskip}
Fitted backgrounds         & $634 \pm 25$          & $4470 \pm 70$~~              \\
\noalign{\smallskip}\hline\noalign{\smallskip}
\text{Fitted } $VV$         & $520 \pm 27$          & ~~$68 \pm 12$              \\
\text{Fitted } \ttbar         & $69 \pm 7$          & $3240 \pm 100$              \\
\text{Fitted single-top}         & $40 \pm 6$          & $1130 \pm 90$~~              \\
\text{Other backgrounds}          & ~$4.8_{-4.8}^{+5.1}$          & $29 \pm 5$              \\
\text{FNP leptons}         & \,$0.02_{-0.02}^{+1.4}$          & \,$0.06_{-0.06}^{+12}$              \\
\noalign{\smallskip}\hline\noalign{\smallskip}
 
\text{Simulated } $VV$  & $376 $          & ~~~~$49 $              \\
\text{Simulated } \ttbar        & ~~$63 $          & $2974 $              \\
\text{Simulated single-top}         & ~~$37 $          & $1040 $             \\
\noalign{\smallskip}\hline\noalign{\smallskip}
\end{tabular*}
\label{tab:CRresults}
\end{table}
\end{DIFnomarkup}

\begin{DIFnomarkup}
\begin{figure}[!htb]
\centering
\begin{subfigure}[b]{0.48\linewidth}
\includegraphics[width=\linewidth]{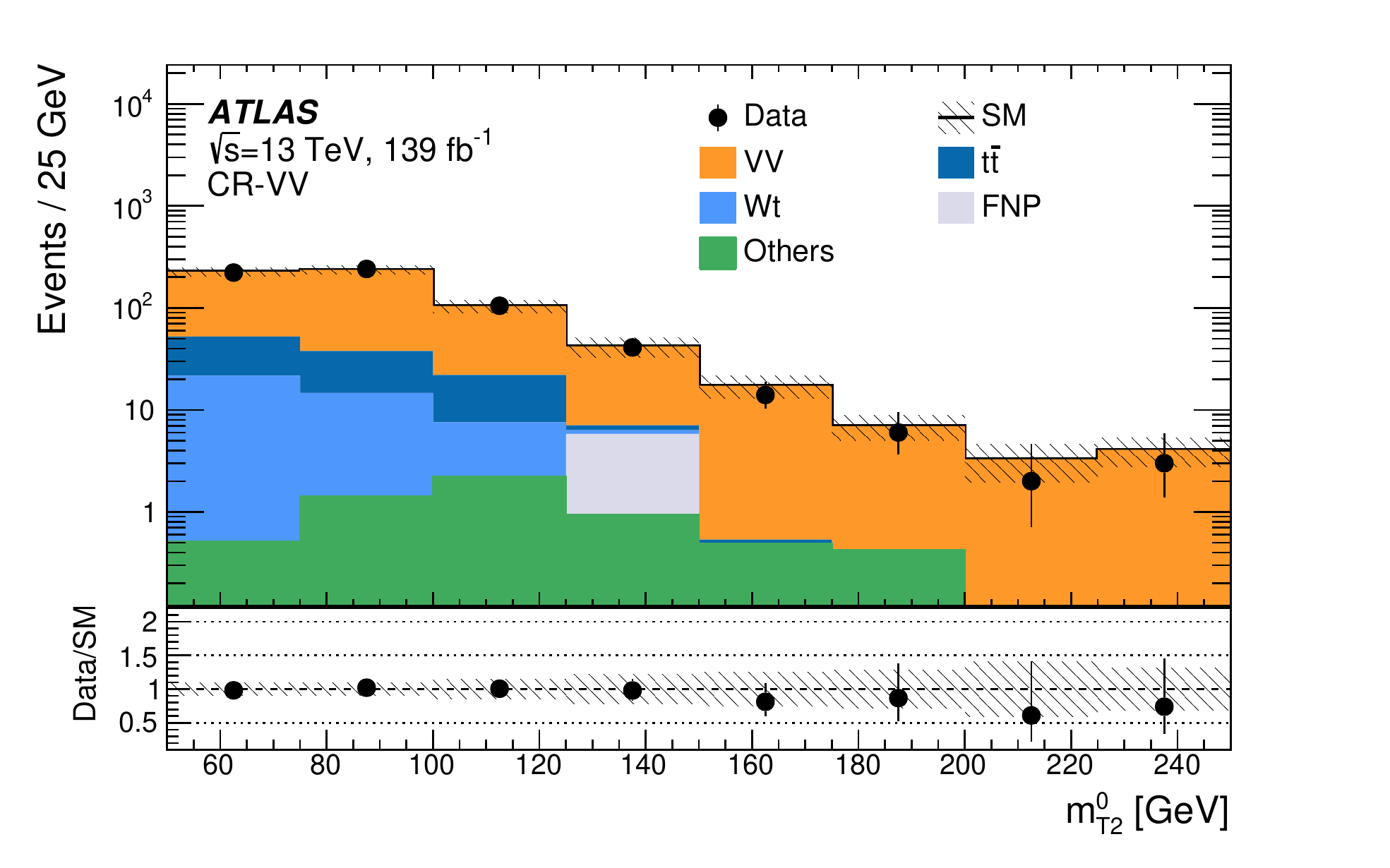}
\caption{$m_{\mathrm{T2}}^{0}$ distribution in CR-VV}
\end{subfigure}
\begin{subfigure}[b]{0.48\linewidth}
\includegraphics[width=\linewidth]{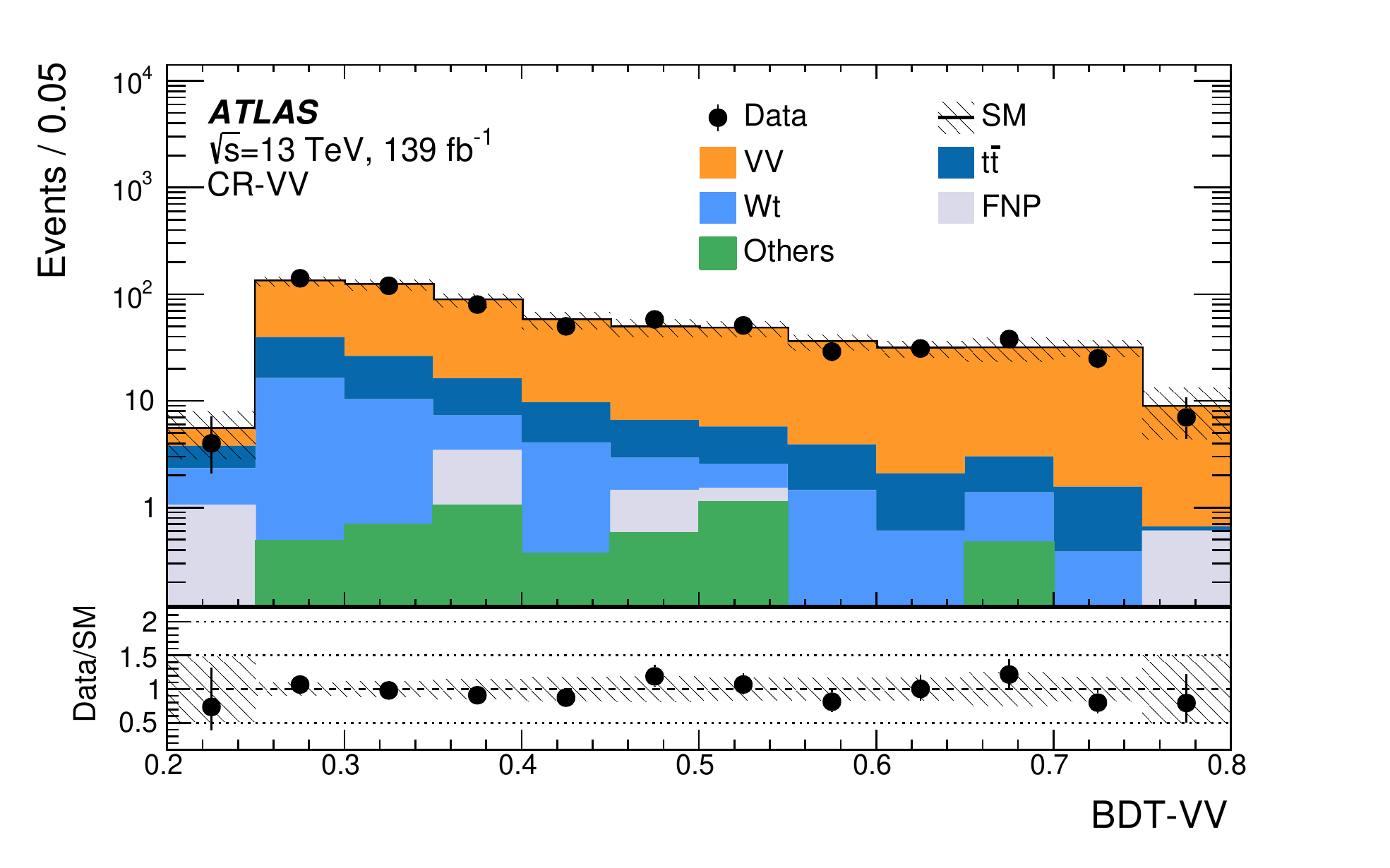}
\caption{BDT-VV distribution in CR-VV}
\end{subfigure}
\begin{subfigure}[b]{0.48\linewidth}
\includegraphics[width=\linewidth]{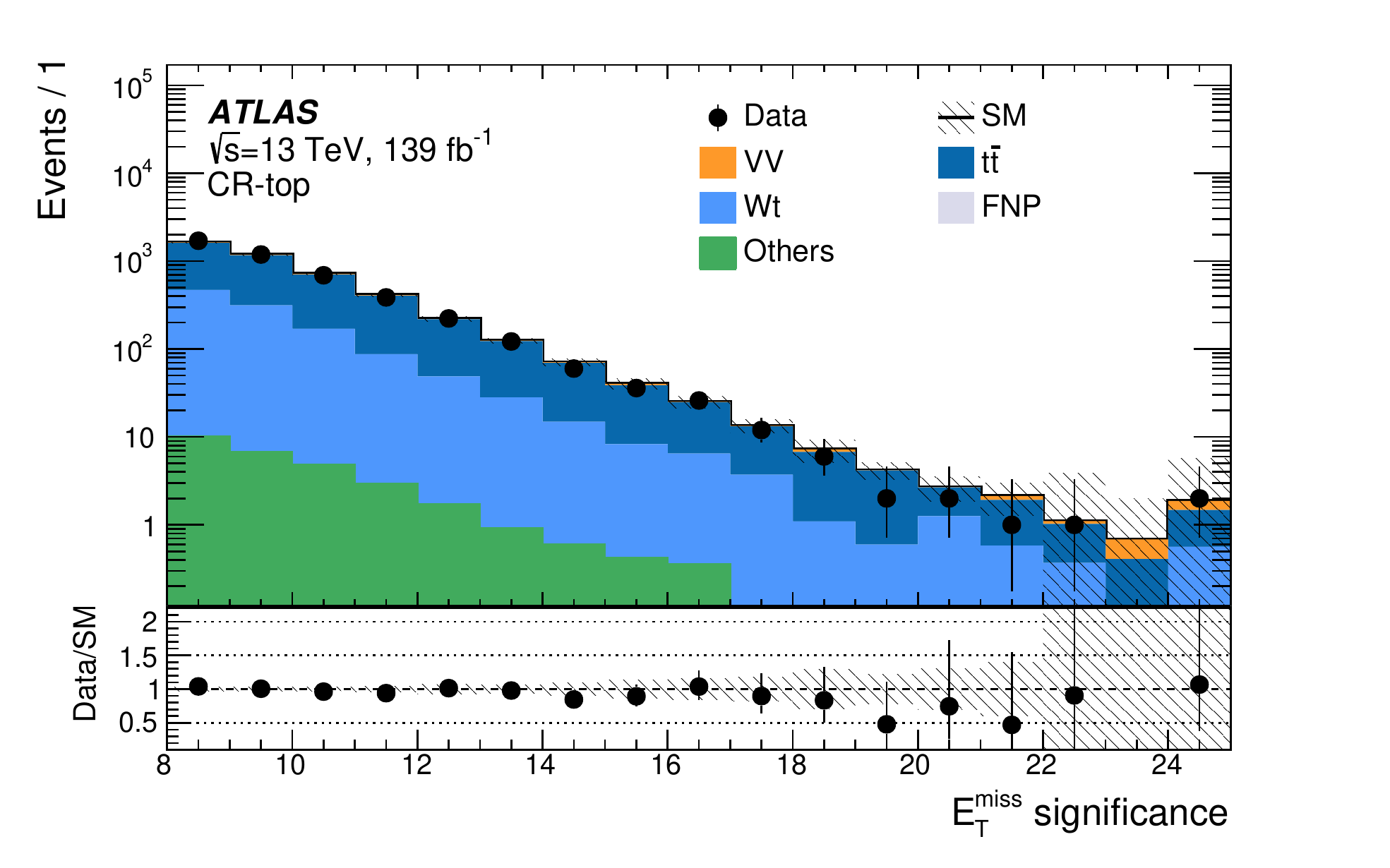}
\caption{\MET\ significance distribution in CR-top}
\end{subfigure}
\begin{subfigure}[b]{0.48\linewidth}
\includegraphics[width=\linewidth]{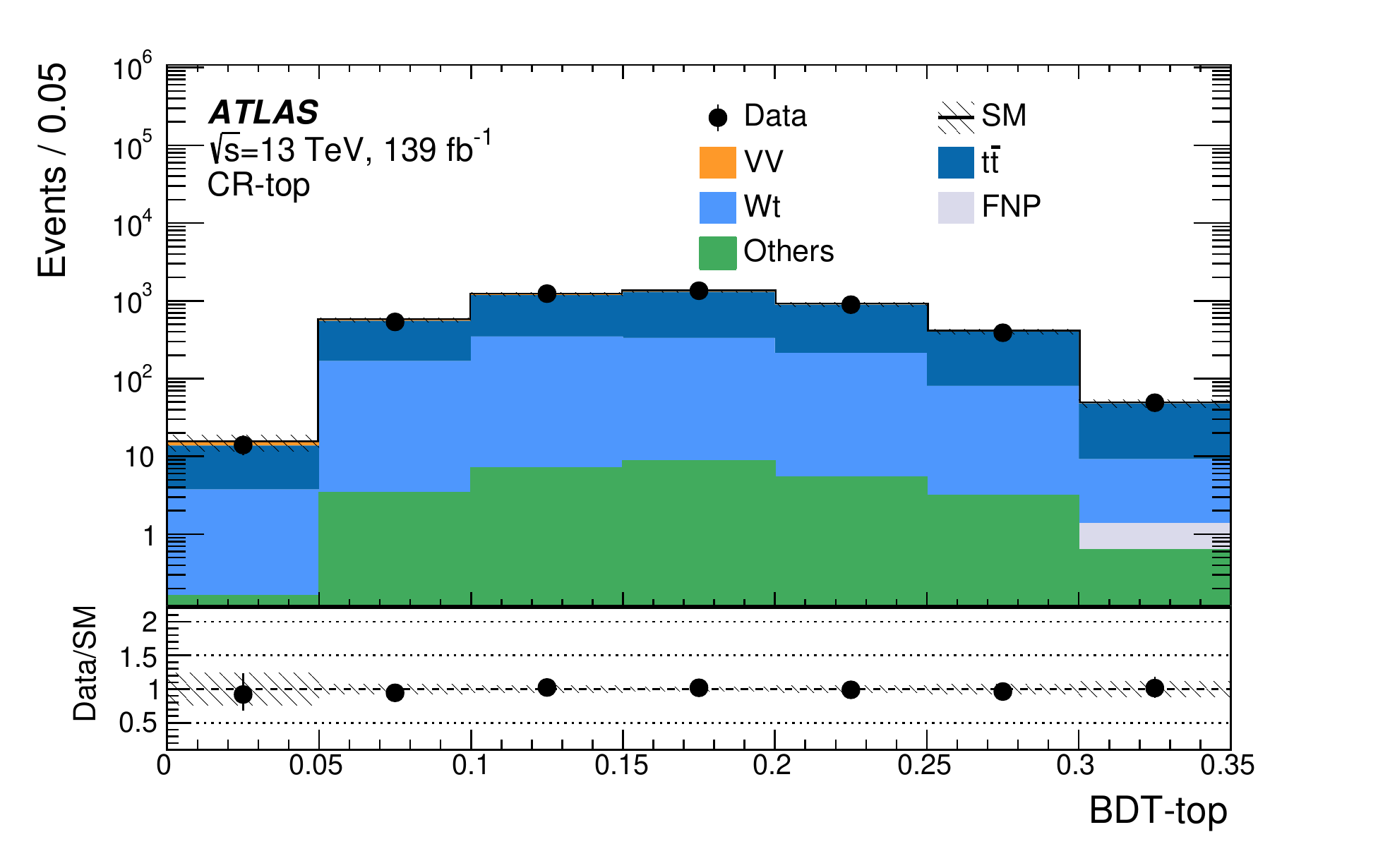}
\caption{BDT-top distribution in CR-top}
\end{subfigure}
 
\caption{ Distributions of (a) $m_{\mathrm{T2}}^{0}$ , (b) BDT-VV in CR-VV, (c) \MET\ significance and (d) BDT-top in CR-top for data and the estimated SM backgrounds. The normalization factors extracted from the corresponding CRs are used to rescale the $t\bar{t}$, single-top-quark and $VV$ backgrounds. The FNP lepton background is calculated using the data-driven matrix method. `Others' include the non-dominant background sources, e.g.\ $t \bar{t}$+$V$, Higgs boson and Drell--Yan events. The uncertainty band includes systematic and statistical errors from all sources and the final bin in each histogram includes the overflow. The lower panels show the ratio of data to the SM background estimate. 
}
\label{fig:CR}
\end{figure}
 
\end{DIFnomarkup}

A set of six validation regions is used to verify that the SM predictions and data agree within uncertainties in regions with a phase-space kinematically close to the SRs, after performing the likelihood fit.
The definitions are reported in the second part of Table~\ref{tab:table_CRVR_charg}.
The regions VR-VV-DF, VR-VV-SF, VR-top-DF and VR-top-SF are designed to have an intermediate BDT-signal selection range compared to the corresponding CRs and SRs.
Regions VR-top0J-DF and VR-top0J-SF are used to validate the extrapolation of the top-quark normalization factor from the region with $n_{ b \text{-tagged jets}}=1$ (CR-top) to regions with $n_{ b \text{-tagged jets}}=0$ (SRs).
Furthermore, VR-top0J-DF and VR-top0J-SF are also used to validate the top-quark background estimate in regions with the same relative fraction of \ttbar and $Wt$ as the one expected in the SRs.
 
\begin{figure}[b]
\centering
\begin{subfigure}[b]{0.47\linewidth}
\includegraphics[width=\linewidth]{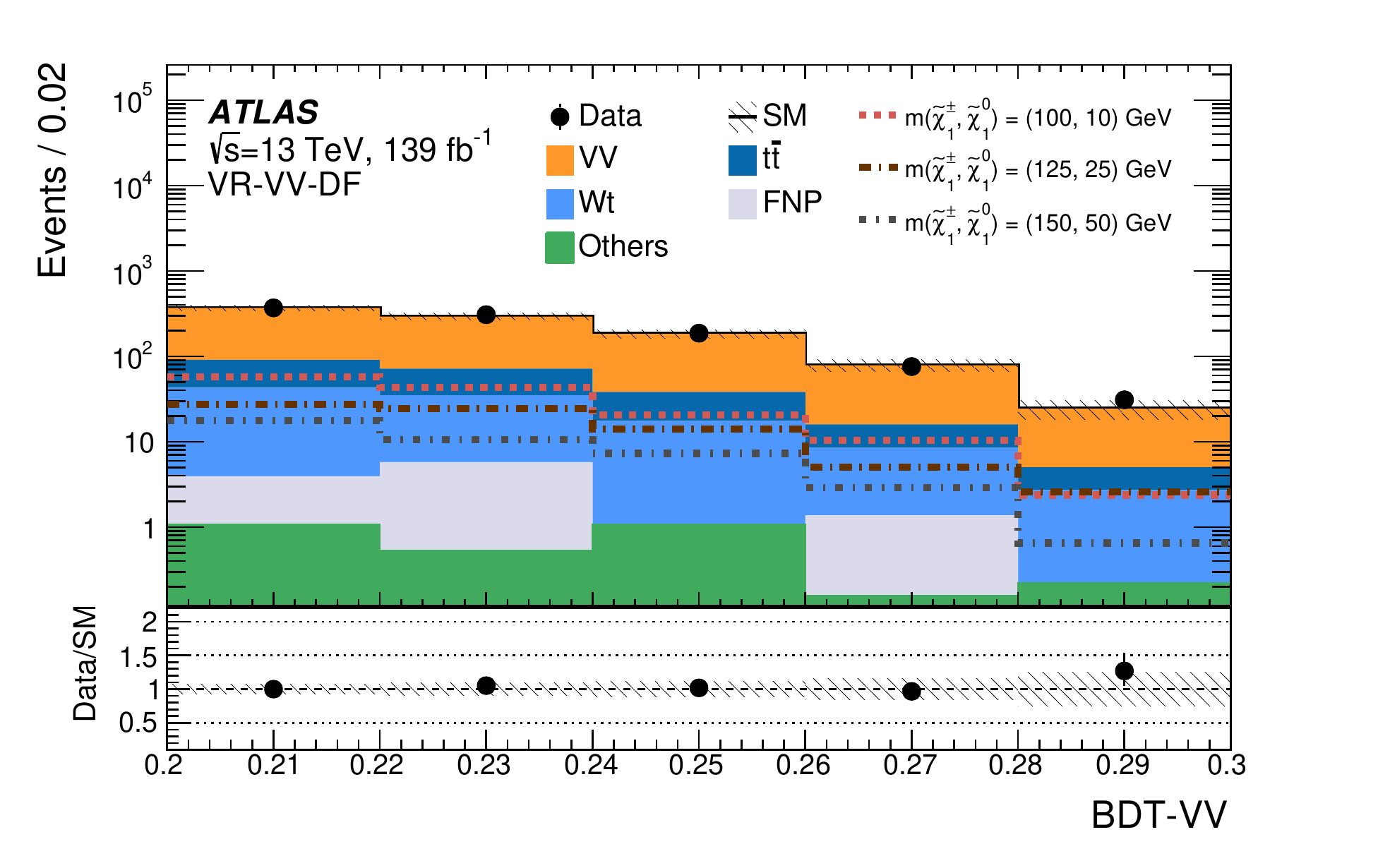}
\caption{BDT-VV distribution in VR-VV-DF}
\end{subfigure}
\begin{subfigure}[b]{0.47\linewidth}
\includegraphics[width=\linewidth]{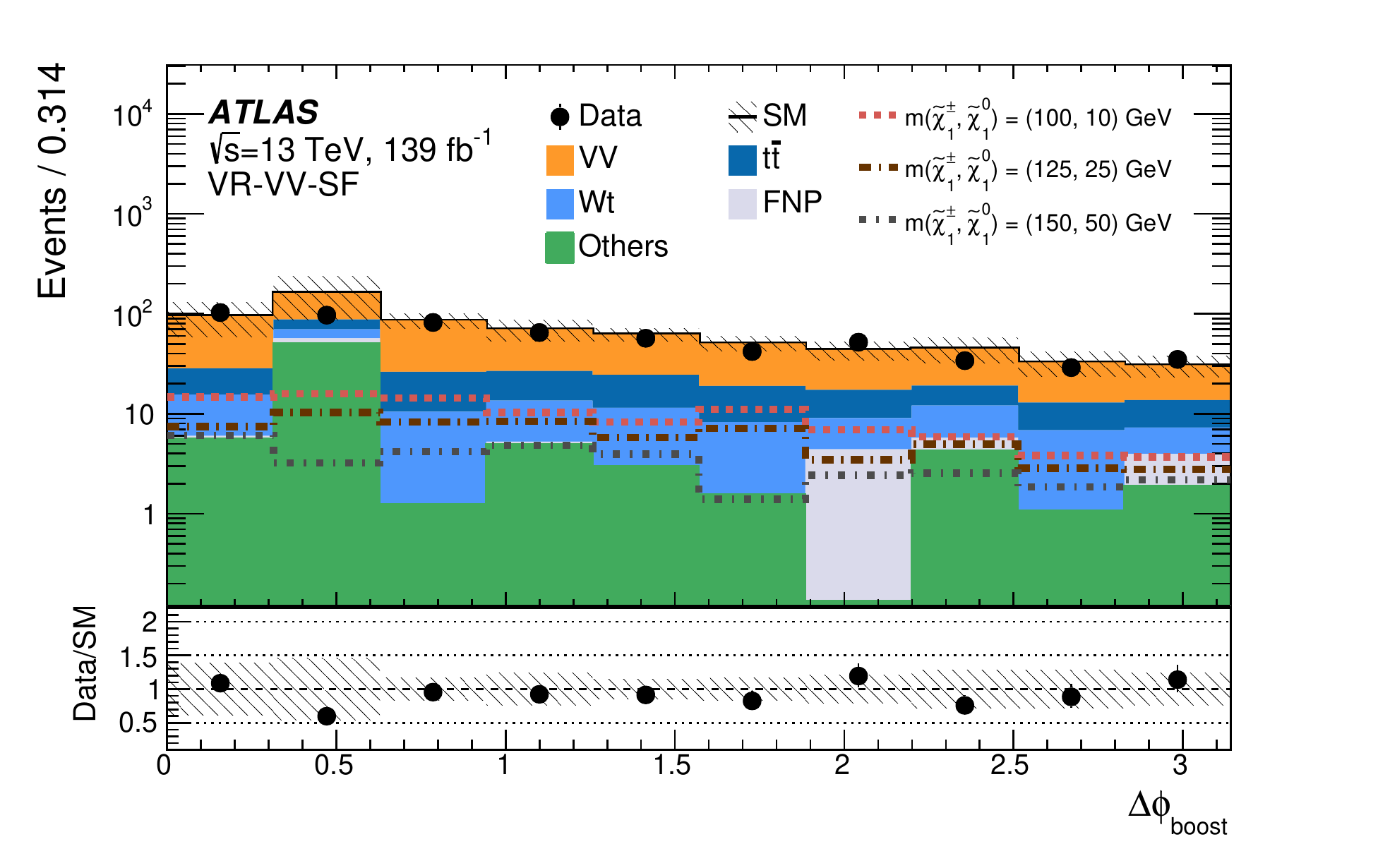}
\caption{ \dphib distribution in VR-VV-SF}
\end{subfigure}
\begin{subfigure}[b]{0.47\linewidth}
\includegraphics[width=\textwidth]{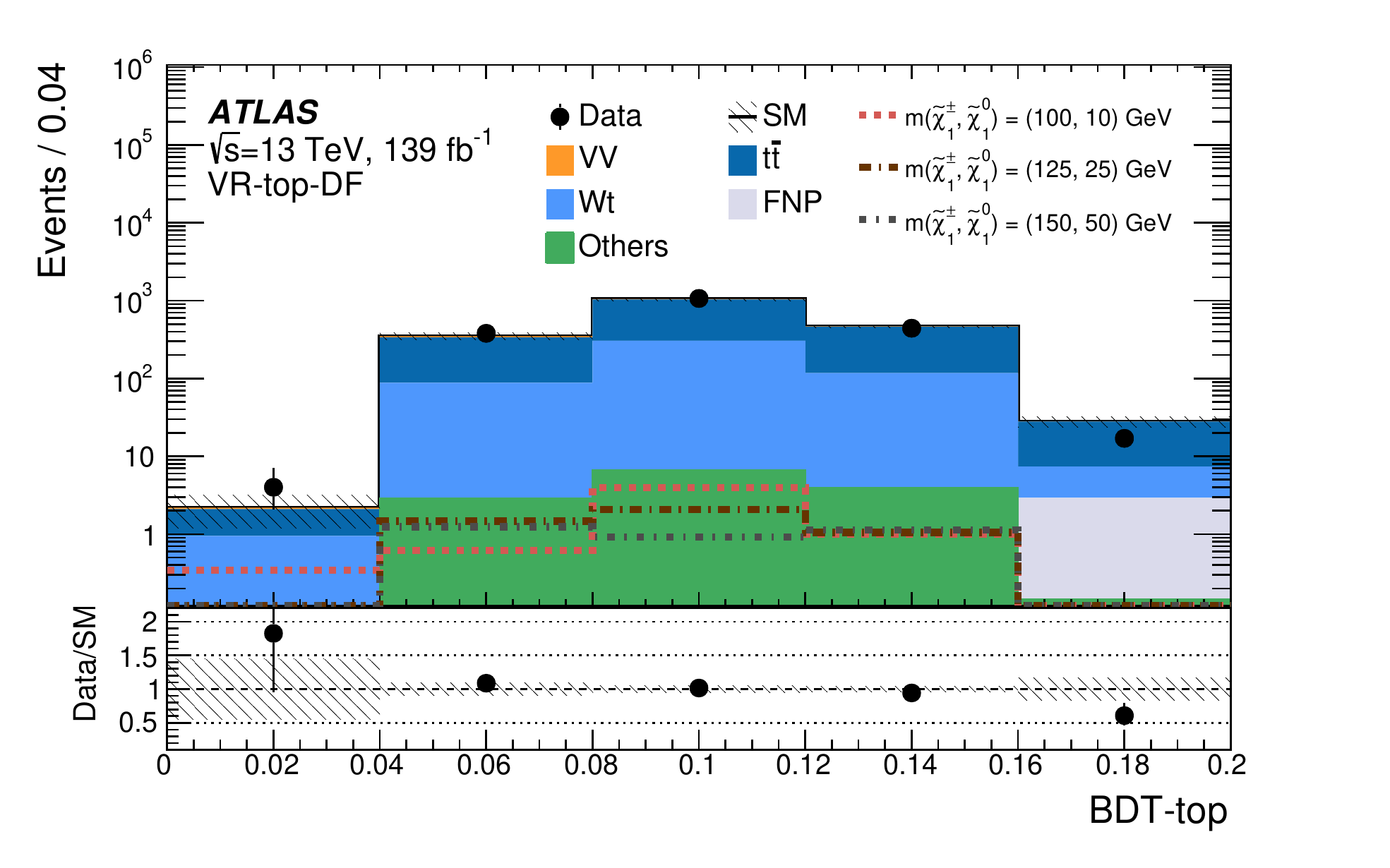}
\caption{BDT-top distribution in VR-top-DF}
\end{subfigure}
\begin{subfigure}[b]{0.47\linewidth}
\includegraphics[width=\linewidth]{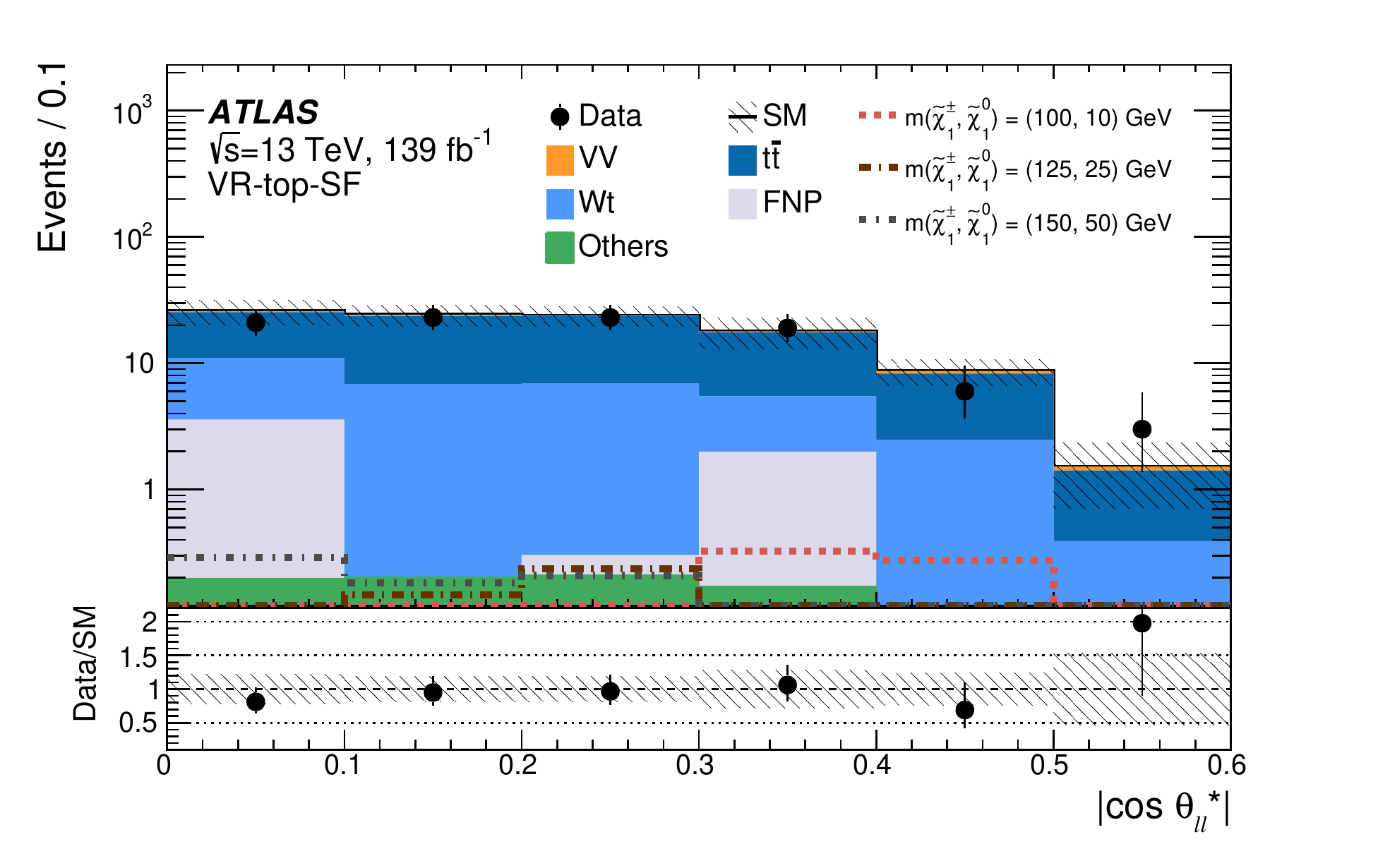}
\caption{$|\cos\theta^{*}_{\ell\ell}|$ distribution in VR-top-SF}
\end{subfigure}
\begin{subfigure}[b]{0.47\linewidth}
\includegraphics[width=\linewidth]{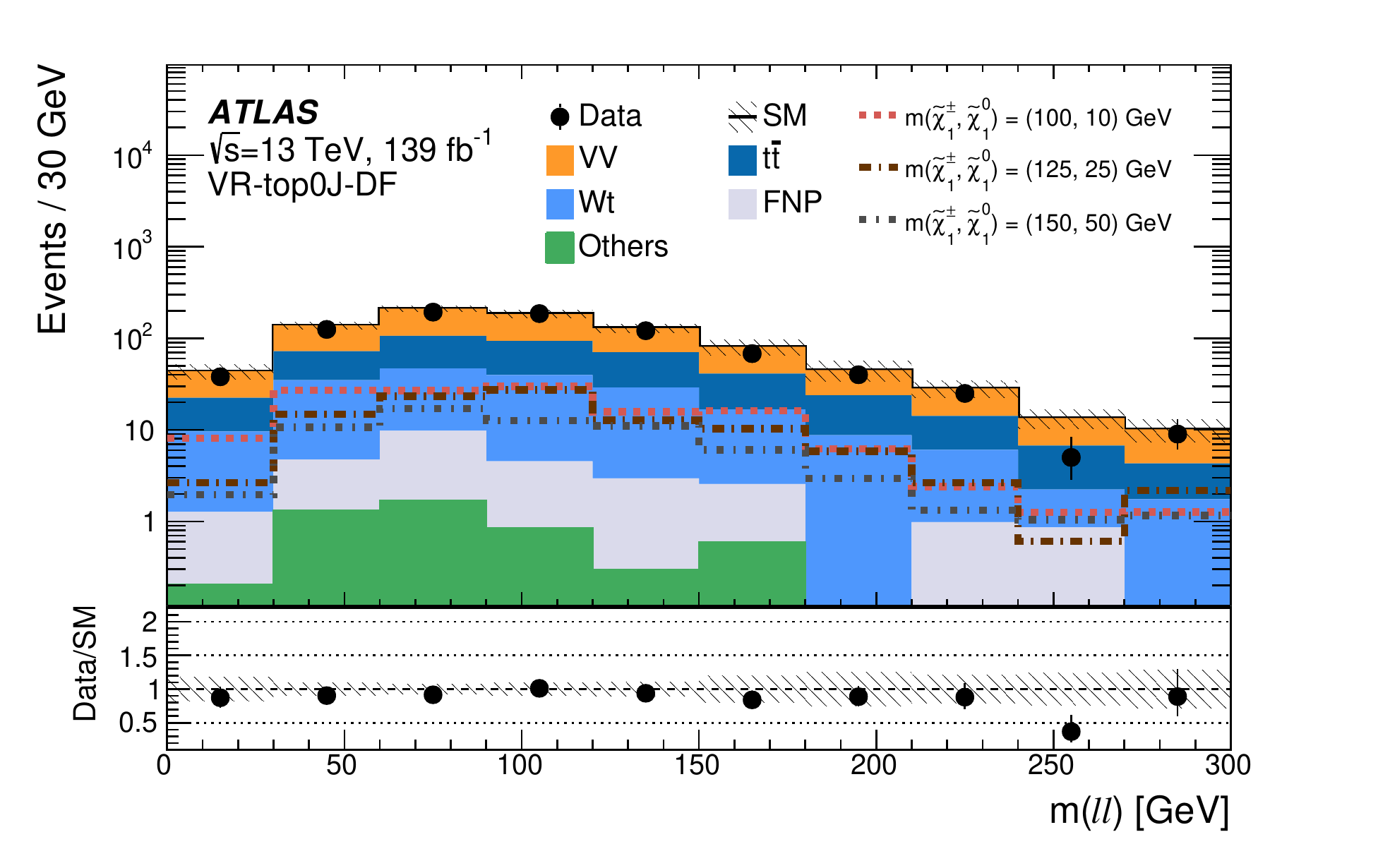}
\caption{\mll distribution in VR-top0J-DF}
\end{subfigure}
\begin{subfigure}[b]{0.47\linewidth}
\includegraphics[width=\linewidth]{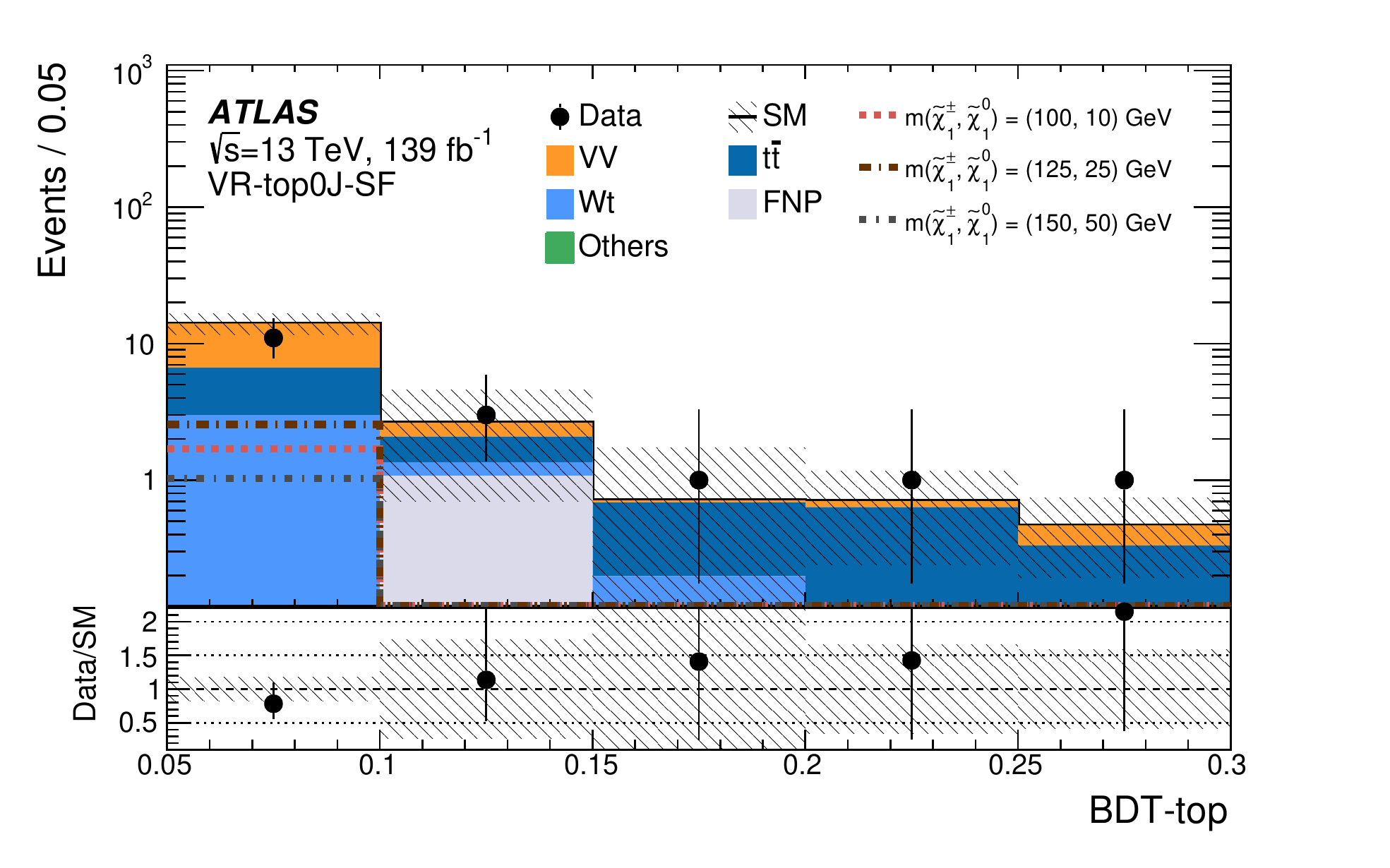}
\caption{BDT-top distribution in VR-top0J-SF}
\end{subfigure}
 
\caption{ Distributions of (a) BDT-VV in VR-VV-DF, (b)  \dphib in VR-VV-SF, (c) BDT-top in VR-top-DF, (d) $|\cos\theta^{*}_{\ell\ell}|$ in VR-top-SF,  (e) \mll in VR-top0J-DF and (f) BDT-top in VR-top0J-SF, for data and the estimated SM backgrounds. The normalization factors extracted from the corresponding CRs are used to rescale the $t\bar{t}$, single-top-quark and $VV$ backgrounds. The FNP lepton background is calculated using the data-driven matrix method. `Others' include the non-dominant background sources, e.g.\ $t \bar{t}$+$V$, Higgs boson and Drell--Yan events. The uncertainty band includes systematic and statistical errors from all sources and the last bin includes the overflow. Distributions for three benchmark signal points are overlaid for comparison. The lower panels show the ratio of data to the SM background estimate.
}
\label{fig:VRdistributions}
\end{figure}

The number of observed events and the predicted background in each VR are shown in Table~\ref{tab:VRresults}. They agree within one standard deviation except in VR-top0J-DF, where a $1.8\sigma$ discrepancy is observed. For backgrounds with a normalization extracted from the likelihood fit, the expected yield from the simulated samples before the likelihood fit is also shown. Figure~\ref{fig:VRdistributions} shows a selection of kinematic distributions for data and the estimated SM background in the validation regions defined in Table~\ref{tab:table_CRVR_charg}. Good agreement is observed in all regions.

\begin{table}
\caption{
Observed event yields and predicted background yields in the VRs for the chargino search. For backgrounds with a normalization extracted from the likelihood fit in the CRs, the expected yield from the simulation before the likelihood fit is also shown. The FNP lepton background is calculated using the data-driven matrix method. `Other backgrounds' include the non-dominant background sources, e.g.\ $t \bar{t}$+$V$, Higgs boson and Drell--Yan events. The uncertainties include both the statistical and systematic contributions.
}
\centering{\scriptsize
\begin{tabular*}{\textwidth}{@{\extracolsep{\fill}}lcccccc}
\noalign{\smallskip}\hline\noalign{\smallskip}
Regions           & VR-VV-DF          & VR-VV-SF          & VR-top-DF                 & VR-top-SF            & VR-top0J-DF           & VR-top0J-SF                 \\[-0.05cm]
\noalign{\smallskip}\hline\noalign{\smallskip}
Observed events          & $972$              & $596$              & $1910$              & $95$              & $810$              & $17$                    \\
\noalign{\smallskip}\hline\noalign{\smallskip}
Fitted backgrounds           & $940 \pm 60$          & $670 \pm 90$          & $1900 \pm 90$~~          & $101 \pm 10$          & $880 \pm 40$          & $18 \pm 4$              \\
\noalign{\smallskip}\hline\noalign{\smallskip}
\text{Fitted } $VV$ 	& $730 \pm 50$          & $400 \pm 50$          & ~~$32 \pm 13$          & ~~$2.2 \pm 2.1$          & $427 \pm 30$          & ~~$8.1 \pm 2.6$              \\
\text{Fitted } \ttbar       & $116\pm 12$          & $111 \pm 11$          & $1350 \pm 50$~~          & $67 \pm 7$          & $260 \pm 21$          & ~~$5.8 \pm 1.8$              \\
\text{Fitted single-top}   & ~~$94 \pm 19$          & ~~$75 \pm 11$          & $500 \pm 60$          & $27 \pm 7$          & $168 \pm 18$          & ~~$4 \pm 1$              \\
\text{Other backgrounds}  & ~~$3.1 \pm 1.5$          & ~~$70 \pm 70$          & $13.6 \pm 2.5$          & ~~$0.8 \pm 0.4$          & ~~$5.2 \pm 1.9$          & ~~$0.05 \pm 0.05$              \\
\text{FNP leptons}      & ~$0.02_{-0.02}^{+2.3}$          & ~~$7 \pm 4$          & ~$0.03_{-0.03}^{+5}$          & ~~$4.2 \pm 1.3$          & $21 \pm 8$          & ~$0.05_{-0.05}^{+0.15}$              \\
\noalign{\smallskip}\hline\noalign{\smallskip}
\text{Simulated } $VV$  & $527 $          & $291 $          & ~~~~$23 $          & ~~~~~\,$1.6 $          & $309$          & ~~~~$5.9 $   \\
\text{Simulated }\ttbar         & $106$          & $102 $          & $1240 $          & $61 $          & $239 $          & ~~~~$5.3 $          \\
\text{Simulated single-top}      & ~~$87 $          & ~~$69 $          & ~~$460 $          & $25 $          & $154 $          & ~~~~$3.2 $            \\
 
\noalign{\smallskip}\hline\noalign{\smallskip}
\end{tabular*}
}
\label{tab:VRresults}
\end{table}

\FloatBarrier
\section{Systematic uncertainties}
\label{sec:Syst}
 
This section describes the uncertainties in the SM background predictions, while the uncertainties associated with the signal models are discussed in Section~\ref{sec:DataMC}.
The likelihood fits used for calculating the results of the two analyses consider all  relevant sources of experimental and theoretical systematic uncertainty affecting the SM background estimates and the signal predictions.
The major sources of uncertainty in the slepton search are related to the FSB estimation,
while in the chargino search the dominant contributions come from the $VV$ theoretical uncertainty, normalization of background processes, and
uncertainty associated with the jet energy scale and resolution and with the $\mathbf p^\mathrm{miss}_\mathrm{T}$ soft-term scale and resolution.
Statistical uncertainties associated with the simulated MC samples are also accounted for. For the chargino search, in the cases where the normalization of background processes ($VV$ and top) are calculated using control regions, the systematic uncertainties only affect the extrapolation to the signal regions.
 
The jet energy scale and resolution uncertainties are calculated as a function of the $p_\mathrm{T}$ and $\eta$ of the jet, and the pile-up conditions and  flavour composition of the selected jet sample. They are derived using a combination of data and simulated samples, through studies including measurements of the transverse momentum balance between a jet and a reference object in dijet, $Z$+jets and $\gamma+$jets events~\cite{PERF-2016-04,JETM-2018-05,PERF-2014-02}.
An additional uncertainty in the modelling of $\mathbf p^\mathrm{miss}_\mathrm{T}$ comes from the soft-term resolution and scale~\cite{PERF-2016-07}. Experimental uncertainties on the scale factors used to account for differences between the data and simulation in $b$-jet identification, lepton reconstruction efficiency and trigger efficiency are also included.
The remaining experimental uncertainties include lepton energy scale and resolution, and are found to be negligible across all analysis regions.
 
Several sources of theoretical uncertainty in the modelling of the dominant backgrounds are considered. Modelling uncertainties affecting diboson, \ttbar, single-top ($Wt$) and $Z$+jets backgrounds are considered in the chargino search, whilst the slepton search only considers modelling uncertainties affecting the $WZ/ZZ$ diboson processes and $Z$+jets, due to the data-driven background estimation method used for the flavour-symmetric backgrounds.
 
The diboson modelling uncertainties are calculated by varying the PDF sets~\cite{Ball:2014uwa} as well as the QCD renormalization and factorization scales used to generate the samples. Uncertainties from missing higher orders are evaluated~\cite{Bothmann:2016nao} using six variations of the QCD factorization and renormalization scales in the matrix elements by factors of $0.5$ and $2$, avoiding variations in opposite directions.
Additional uncertainties in the resummation and matching scales between the matrix element generator and parton shower are considered.
 
The \ttbar\ background estimate is affected by modelling uncertainties associated with the parton shower modelling, the different approaches commonly used in the matching between the matrix element and the parton shower, and the modelling of initial- and final-state radiation (ISR/FSR). Uncertainties in the parton shower simulation are estimated by comparing samples generated with \POWHEGBOX interfaced to either \PYTHIA[8.186] or \HERWIG[7.04]~\cite{Bahr:2008pv,Bellm:2015jjp}. The ISR/FSR uncertainties are calculated by comparing the predictions of the nominal sample with alternative scenarios with the relevant generator parameters varied~\cite{ATL-PHYS-PUB-2017-007}. The uncertainty associated with the choice of event generator is estimated by comparing the nominal samples with samples generated with \AMCatNLO interfaced to \PYTHIA[8.186]~\cite{ATL-PHYS-PUB-2017-007}. Finally, for single-top-quark production, an uncertainty is assigned to the treatment of the interference between the $Wt$ and $t\bar{t}$ samples. This is done by comparing the nominal sample generated using the diagram removal method with a sample generated using the diagram subtraction method~\cite{Frixione:2008yi,ATL-PHYS-PUB-2016-020}.
 
The $Z$+jets background estimate is affected by QCD factorization and renormalization scale uncertainties. Uncertainties in the resummation and matching scales between the matrix element generator and parton shower are also considered.
 
The uncertainty in the MM estimate of the FNP background has several components. The `real' efficiencies and the electron light-flavour fake rate (which are calculated using MC simulation) are affected by the experimental uncertainties of the scale factors applied to account for lepton trigger, identification, reconstruction and isolation efficiency differences between data and simulation. For the heavy-flavour fake rate, the effects of uncertainties in the subtraction of the prompt-lepton contamination in the control region are calculated by varying this contamination and evaluating the changes in the FNP background estimates.
Finally, uncertainties in the expected composition of the FNP leptons in the signal regions are included, along with the statistical uncertainties of the `real' efficiencies and fake rates used in the calculation.
 
For the slepton search, additional uncertainties associated with the data-driven background estimate of the flavour-symmetric backgrounds (FSB estimate) discussed in Section~\ref{sec:backgrounds} are also applied. The statistical uncertainty of the DF sample is included.  Uncertainties in the $\kappa$ and $\alpha$ factors that account for the reconstruction, identification and trigger efficiency differences between muons and electrons are obtained by considering the differences between global efficiencies calculated in data and simulation. Finally, additional uncertainties are applied to account for possible changes in the results if the events are reweighted as a function of the sub-leading lepton $p_\mathrm{T}$ instead of the leading lepton $p_\mathrm{T}$, and for the choice of fitting function for the dependence of $\kappa$ on this variable. A summary of the impact of the systematic uncertainties on the background yields in the inclusive SRs requiring $m_{\mathrm{T2}}^{100}\in [100,\infty$)~\GeV, after performing the likelihood fit, is shown in Table~\ref{tab:systematicSummary_slep} for the 0J and 1J selections.
 
\begin{table}
\caption{Breakdown of the dominant systematic uncertainties in background estimates in the inclusive SRs requiring $m_{\mathrm{T2}}^{100}\in [100,\infty$)~\GeV\ for the 0J and 1J selections in the slepton search.
The individual uncertainties can be correlated, and do not necessarily sum in quadrature to the total background uncertainty.
The percentages show the size of the uncertainty relative to the total expected background.
\label{tab:systematicSummary_slep}}
\begin{center}
\setlength{\tabcolsep}{0.0pc}
\begin{tabular}{p{0.5\linewidth}p{0.16\linewidth}p{0.16\linewidth}}
\noalign{\smallskip}\hline\noalign{\smallskip}{Region}            & \hspace{-10pt} SR-0J                 & \hspace{-10pt} SR-1J               \\
$m_{\mathrm{T2}}^{100}$ [\GeV]     & \hspace{-10pt} $\in$[100,$\infty$)   & \hspace{-10pt} $\in$[100,$\infty$) \\
\noalign{\smallskip}\hline\noalign{\smallskip}
\end{tabular}
\begin{tabular}{p{0.5\linewidth}p{0.16\linewidth}p{0.16\linewidth}}
Total background expectation  & \hspace{1.7pt} 76    & \hspace{1.7pt} 78   \\
\noalign{\smallskip}\hline\noalign{\smallskip}{$\!\begin{aligned}
& \text{MC and FSB statistical uncertainties }                                 \\[-2pt]
& \text{FSB estimate}                                             \\[-2pt]
& \text{FNP leptons}   \\[-2pt]
& Z/\gamma^{*} (\rightarrow \ell\ell) \text{+jets theoretical uncertainties} \\[-2pt]
& \MET \text{ modelling}                                               \\[-2pt]
& \text{Jet energy scale}                                              \\[-2pt]
& \text{Jet energy resolution}                                         \\[-2pt]
& b\text{-tagging}                                                     \\[-2pt]
& \text{Lepton modelling}
\end{aligned}
$}
&   {$\!\begin{aligned}
14   &\%   \\[-2pt]
9    &\%   \\[-2pt]
5    &\% \\[-2pt]
<1   &\%   \\[-2pt]
2.3  &\%   \\[-2pt]
<1   &\%   \\[-2pt]
<1   &\%   \\[-2pt]
<1   &\%   \\[-2pt]
1    &\%
\end{aligned}
$}
&   {$\!\begin{aligned}
13   &\%   \\[-2pt]
9    &\%   \\[-2pt]
4    &\% \\[-2pt]
3    &\%   \\[-2pt]
<1   &\%   \\[-2pt]
<1   &\%   \\[-2pt]
1    &\%   \\[-2pt]
<1   &\%   \\[-2pt]
<1   &\%
\end{aligned}
$} \\
\noalign{\smallskip}\hline\noalign{\smallskip}{$\!\begin{aligned}
& \text{Total systematic uncertainty}
\end{aligned}
$}
&   {$\!\begin{aligned}
\hspace{5pt} 17 &\%
\end{aligned}
$}
&   {$\!\begin{aligned}
\hspace{5pt} 17 &\%
\end{aligned}
$} \\
\noalign{\smallskip}\hline\noalign{\smallskip}
\end{tabular}
\end{center}
\end{table}

For the chargino search, a summary of the impact of the systematic uncertainties on the background yields in the inclusive region SR$^{\text{-DF BDT-signal}\in(0.81,1]}_{\text{-SF BDT-signal}\in(0.77,1]}$, obtained as a combination of the integrals of all DF and SF binned regions in Table~\ref{tab:SRchargdefs}, is shown in Table~\ref{tab:systematicSummary_charg} after performing the likelihood fit.

\begin{table}
\caption{Breakdown of the dominant systematic uncertainties in background estimates in the inclusive region SR$^{\text{-DF BDT-signal}\in(0.81,1]}_{\text{-SF BDT-signal}\in(0.77,1]}$ for the chargino search.
The individual uncertainties can be correlated, and do not necessarily sum in quadrature to
the total background uncertainty. The percentages show the size of the uncertainty relative to the total expected background.}
\label{tab:systematicSummary_charg}
\begin{center}
\setlength{\tabcolsep}{0.0pc}
\begin{tabular}{p{0.5\linewidth}p{0.26\linewidth}}
\noalign{\smallskip}\hline\noalign{\smallskip}{Region}                            & \hspace{-45pt} SR$^{\text{-DF BDT-signal}\in(0.81,1]}_{\text{-SF BDT-signal}\in(0.77,1]}$ \\
\noalign{\smallskip}\hline\noalign{\smallskip}
\end{tabular}
\begin{tabular}{p{0.5\linewidth}p{0.26\linewidth}}
Total background expectation      & \hspace{1.7pt}  630  \\
\noalign{\smallskip}\hline\noalign{\smallskip}{$\!\begin{aligned}
& \MET \text{ modelling}                                 \\[-2pt]
& \text{Diboson theoretical uncertainties}                                 \\[-2pt]
& \text{Jet energy scale}                                 \\[-2pt]
& VV\text{ normalization}                                 \\[-2pt]
& \text{Jet energy resolution}                                 \\[-2pt]
& \text{MC statistical uncertainties}                                 \\[-2pt]
& \text{Lepton modelling}                                 \\[-2pt]
& \text{Top theoretical uncertainties}                                 \\[-2pt]
& \ttbar \text{ normalization}                                 \\[-2pt]
& \text{FNP leptons}                                 \\[-2pt]
& b\text{-tagging}                                 \\[-2pt]
& \text{$Z/\gamma^*(\rightarrow \ell\ell )$+jets theoretical uncertainties}
\end{aligned}
$}
&   {$\!\begin{aligned}
9.1    &\%   \\[-2pt]
5.8    &\%   \\[-2pt]
5.2     &\%   \\[-2pt]
3.6     &\%   \\[-2pt]
1.7    &\%   \\[-2pt]
1.7    &\%   \\[-2pt]
1.2    &\%   \\[-2pt]
1&\%   \\[-2pt]
1    &\%   \\[-2pt]
0.8    &\%   \\[-2pt]
0.7    &\%   \\[-2pt]
0.04   &\%
\end{aligned}
$} \\
\noalign{\smallskip}\hline\noalign{\smallskip}{$\!\begin{aligned}
& \text{Total systematic uncertainty}
\end{aligned}
$}
&   {$\!\begin{aligned}
\hspace{5pt} 12  &\%
\end{aligned}
$} \\
\noalign{\smallskip}\hline\noalign{\smallskip}
\end{tabular}
\end{center}
\end{table}

\section{Results}
\label{sec:result}

The results of the two searches are interpreted in the context of the slepton and chargino simplified models shown
in Figure~\ref{fig:c1c1viawwfeynman}, and as general limits on new-physics cross-sections.
 
The statistical interpretation of the results is performed using the HistFitter~\cite{HistFitter} framework.
The likelihood is a product of Poisson probability density functions, describing the observed number of events in each CR/SR, and Gaussian distributions that describe the nuisance parameters associated with each of the systematic uncertainties.
Furthermore, Poisson distributions are used for MC statistical uncertainties. Systematic uncertainties that are correlated between different samples are accounted for in the fit configuration by using the same nuisance parameter.
In particular, experimental systematic uncertainties are correlated between background and signal samples for all regions.
The uncertainties are applied in each of the CRs and SRs and their effect is correlated for events across all regions in the fit.
 
The background fit strategy differs between the two searches.
The chargino search uses data in the CRs and the likelihood fit is performed to constrain the nuisance parameters of
the likelihood function, which include the background normalization factors and parameters associated
with the systematic uncertainties. The slepton search uses the FSB prediction in the SRs,
and the likelihood fit is used to constrain the nuisance parameters associated with the systematic uncertainties.
In both cases, the results of the background fit are used to test the compatibility of the observed
data and the background estimates in the inclusive SRs.
 
The CL$_{\text{s}}$ method~\cite{Read:2002hq} is used to set model-independent upper limits at 95\% confidence level (CL) on the visible signal cross-section $\sigma^\mathrm{obs}$, defined as the cross-section times acceptance times efficiency, for processes beyond the SM. They are derived in each inclusive SR by performing a fit that includes the CRs, the observed yield in the SR as a constraint, and a signal yield in the SR as a free parameter of interest.
The observed ($S^{0.95}_\mathrm{obs}$) and expected ($S^{0.95}_\mathrm{exp}$) limits at 95\% CL on the numbers of events from processes beyond the SM in the inclusive SRs are calculated.
The $p_0$-values, which represent the probability of the SM background alone to fluctuate to the observed number of events or higher, are also included in the results and are capped at $p_0 = 0.50$.
 
Exclusion limits at 95\% CL are placed on the masses of the sleptons, chargino and neutralino. The CL$_{\text{s}}$ prescription is also used in this
case, including the data in the binned SRs in the simultaneous likelihood fit.

\subsection{Results of the slepton search}

The predicted number of background events obtained by applying the efficiency correction method to compute the expected number of FSB events, together with the observed data in the binned SRs defined in Table~\ref{tab:SRslepdefs}, are shown in Figure~\ref{fig:SRpull_slep} for 0J  and 1J selections.
In the binned SR-0J, the expected background exceeds the observed data in two $\mTtwoh$ bins, with a local significance of about $2\sigma$. The same behaviour is observed in these bins when using pure MC simulations to estimate the background, so the disagreement most likely arises from statistical fluctuations in data.
In the binned SR-1J, there are excesses of data of about $1.5\sigma$ in two $\mTtwoh$ bins, while the expected background exceeds the observed data with a local significance of $3.5\sigma$ in one $\mTtwoh$ bin. These discrepancies are found to be strictly correlated with statistical fluctuations in the distribution of DF events in data which are used to estimate the FSB. This is observed when comparing pure MC simulations with DF data in the SRs. Furthermore, when comparing pure MC simulations with SF data in the SRs, fluctuations of the data in the opposite direction are observed. The combination of the two effects enhances the discrepancy.
 
The observed and predicted numbers of background events in the inclusive SRs are reported in Table~\ref{tab:SRinc_slep},
together with the model-independent upper limits on the visible signal cross-section $\sigma^\mathrm{obs}$,
the observed and expected limits at 95\% CL on the number of potential beyond-the-SM events, and the $p_0$-values.
Exclusion limits at 95\% CL on the masses of the sleptons and neutralino are shown in Figure~\ref{fig:exclusion_slep} for mass-degenerate $\tilde{e}_{\textup{L,R}}/\tilde{\mu}_{\textup{L,R}}$,
bridging  the gap between previous ATLAS searches and surpassing limits from LEP: sleptons up to 150~\GeV\ are excluded at 95\% CL in the case of a 50~\GeV\ mass-splitting between the sleptons and the LSP.
 
Exclusion limits are also set for selectrons and smuons separately, considering the same selection (including both dielectron and dimuon events in the likelihood fit) used for the general result.
These are shown in Figure~\ref{fig:sleptonExclusions} for single slepton species
$\tilde{e}_{\textup{R}}$, $\tilde{e}_{\textup{L}}$ and $\tilde{\mu}_{\textup{L}}$ along with combined limits for mass-degenerate $\tilde{e}_{\textup{L,R}}$ and $\tilde{\mu}_{\textup{L,R}}$.
Concerning this last case, parts of the region excluded by this search in the $m(\tilde{\mu})-m(\tilde{\chi}_1^0)$ plane are compatible with the $(g-2)_\mu$ anomaly for small $\tan\beta$ values~\cite{ATL-PHYS-PUB-2022-013}.

\begin{DIFnomarkup}
\begin{figure}[h]
\centering
\includegraphics[width=\linewidth]{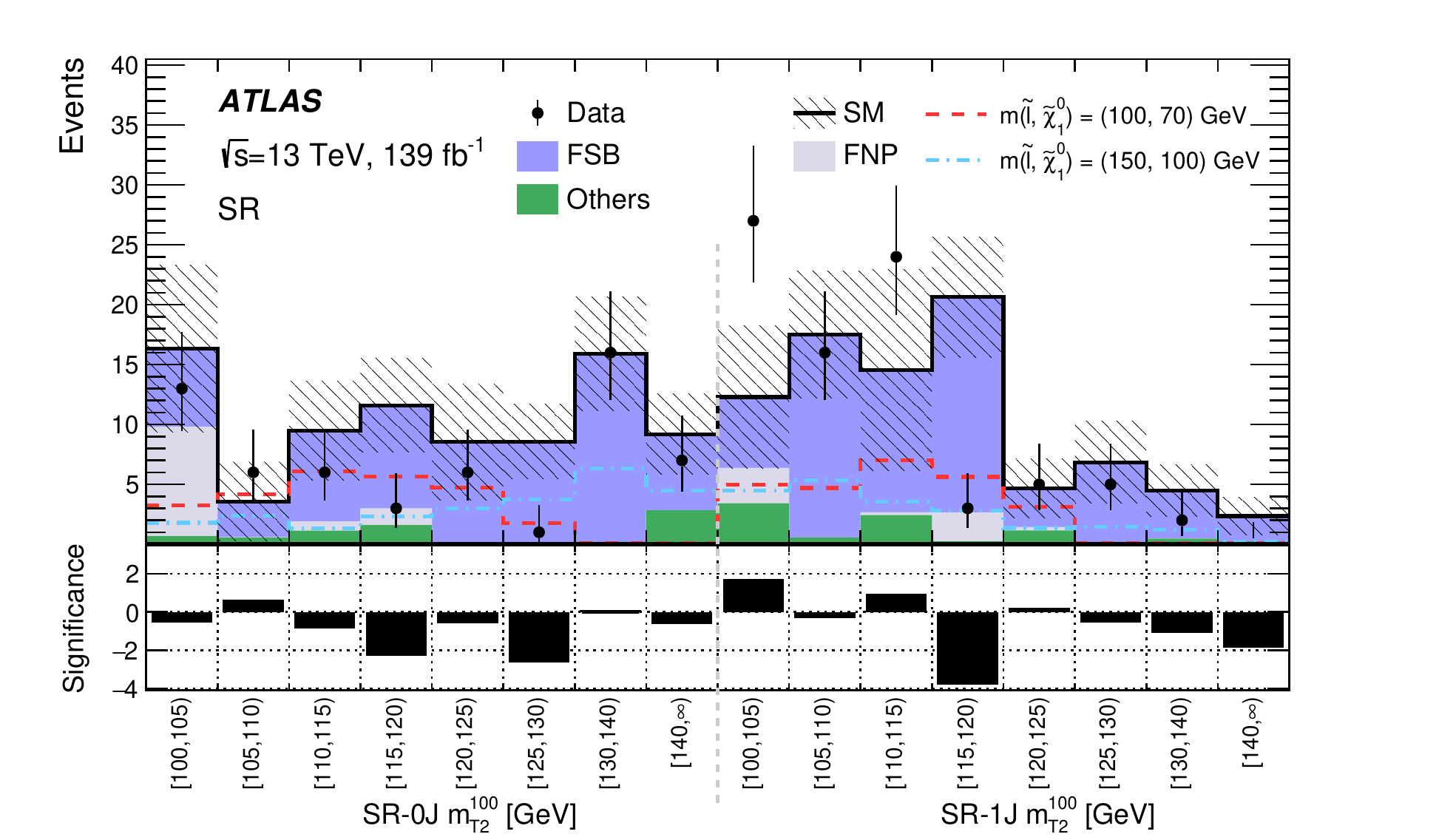}
\caption{The upper panel shows the observed number of events in each of the binned SRs defined in Table~\ref{tab:SRslepdefs}, together with the expected SM backgrounds obtained after applying the efficiency correction method to compute the number of expected FSB events. `Others' include the non-dominant background sources, e.g.\ $t \bar{t}$+$V$, Higgs boson and Drell--Yan events. The uncertainty band includes systematic and statistical errors from all sources.
The distributions of two signal points with mass splittings $\Delta m(\tilde{\ell},\tilde{\chi}_1^0) = m(\tilde{\ell})-m(\tilde{\chi}_1^0)$ of 30~\GeV\ and 50~\GeV\ are overlaid.
The lower panel shows the significance as defined in Ref.~\cite{Cousin:2008clt}. }
\label{fig:SRpull_slep}
\end{figure}
\end{DIFnomarkup}

\begin{DIFnomarkup}
\begin{table}
\centering
\setlength{\tabcolsep}{0.0pc}
\caption{Observed event yields and predicted background yields for the inclusive SRs defined in Table~\ref{tab:SRslepdefs} are reported in the left part of the table, as obtained using the background fit described in Section~\ref{sec:result}. The right part shows the model-independent upper limits at 95\% CL on the observed and expected number of beyond-the-SM events $S^{0.95}_\mathrm{obs/exp}$ and on the effective beyond-the-SM cross-section $\sigma^\mathrm{obs}$ ($\langle\mathrm{A}\epsilon\sigma\rangle^{0.95}_{\mathrm{obs}}$). They were obtained using a fit which includes the SRs, also described in Section~\ref{sec:result}. The $\pm 1 \sigma$ variations of $S^{0.95}_\mathrm{exp}$ are provided. The last column shows the $p_0$-value of the SM-only hypothesis.
For SRs where the data yield is smaller than expected, the $p_0$-value is capped at 0.50.}
\bgroup
\def\arraystretch{1.5}
\begin{tabular*}{\textwidth}{@{\extracolsep{\fill}}lcc@{\hskip 0.3in} | cccc}  
\noalign{\smallskip}\hline\noalign{\smallskip}{\textbf{ Signal region }[\GeV]}    & Observed & Expected     & $\sigma^\mathrm{obs}$ [fb]  &  $S_{\textrm{obs}}^{0.95}$  & $S_{\textrm{exp}}^{0.95}$  & $p_0$  \\
\noalign{\smallskip}\hline\noalign{\smallskip}
$\textrm{SR-0J}\ {\mTtwoh \in [100,\infty )}$ & $58$ & $76 \pm 13$  & $0.13$ &  $18.3$ & $ ~{ 26 }^{ +10 }_{ -7 }$ & $ 0.50$ \\%
$\textrm{SR-0J}\ {\mTtwoh \in [110,\infty )}$  & $39$ & $58 \pm 11$  & $0.09$ &  $13.2$ & $ { 21 }^{ +8 }_{ -6 }$  & $ 0.50$ \\%
$\textrm{SR-0J}\ {\mTtwoh \in [120,\infty )}$ & $30$ & $40 \pm 8$~~ &  $0.10$ &  $13.5$ & $ { 18 }^{ +7 }_{ -5 }$  & $ 0.50$ \\%
$\textrm{SR-0J}\ {\mTtwoh \in [130,\infty )}$ & $23$ & $24 \pm 6$~~ & $0.10$ &  $14.2$ & $ { 15}^{ +6 }_{ -4 }$  & $ 0.50$ \\%
$\textrm{SR-0J}\ {\mTtwoh \in [140,\infty )}$ & ~~$7$ & $9.2 \pm 3.4$  & $0.05$ &  ~~$7.5$ & ~$ { 8.6 }^{ +4}_{ -2.5 }$  & $ 0.50$ \\%
$\textrm{SR-1J}\ {\mTtwoh \in [100,\infty )}$ & $82$ & $78 \pm 13$ & $0.24$ &  $33.5$ & ~$ { 31 }^{ +11}_{ -8 }$ & $ 0.41$ \\%
$\textrm{SR-1J}\ {\mTtwoh \in [110,\infty )}$ & $39$ & $50 \pm 17$ & $0.17$ &  $24.0$ & $ { 28}^{ +9 }_{ -7 }$  & $ 0.50$ \\%
$\textrm{SR-1J}\ {\mTtwoh \in [120,\infty )}$  & $12$ & $16 \pm 5$~~  & $0.07$ &  ~~$9.5$ & $ { 12 }^{ +5 }_{ -3 }$  & $ 0.50$ \\%
$\textrm{SR-1J}\ {\mTtwoh \in [130,\infty )}$  & ~~$2$ & $6.9 \pm 2.8$  & $0.03$ &  ~~$3.9$ & ~$ { 6.1 }^{ +3.0}_{ -1.9 }$  & $ 0.50$ \\%
$\textrm{SR-1J}\ {\mTtwoh \in [140,\infty )}$ & ~~$0$ & $2.4 \pm 1.6$  & $0.02$ &  ~~$2.4$ & ~$ { 3.4 }^{ +2.2 }_{ -1.2 }$  & $ 0.50$ \\%
\noalign{\smallskip}\hline\noalign{\smallskip}
\end{tabular*}
\egroup
\label{tab:SRinc_slep}
\end{table}
\end{DIFnomarkup}

\begin{DIFnomarkup}
\begin{figure}[htb!]
\centering
\begin{subfigure}[a]{\linewidth}
\centering
\includegraphics[width=0.7\linewidth]{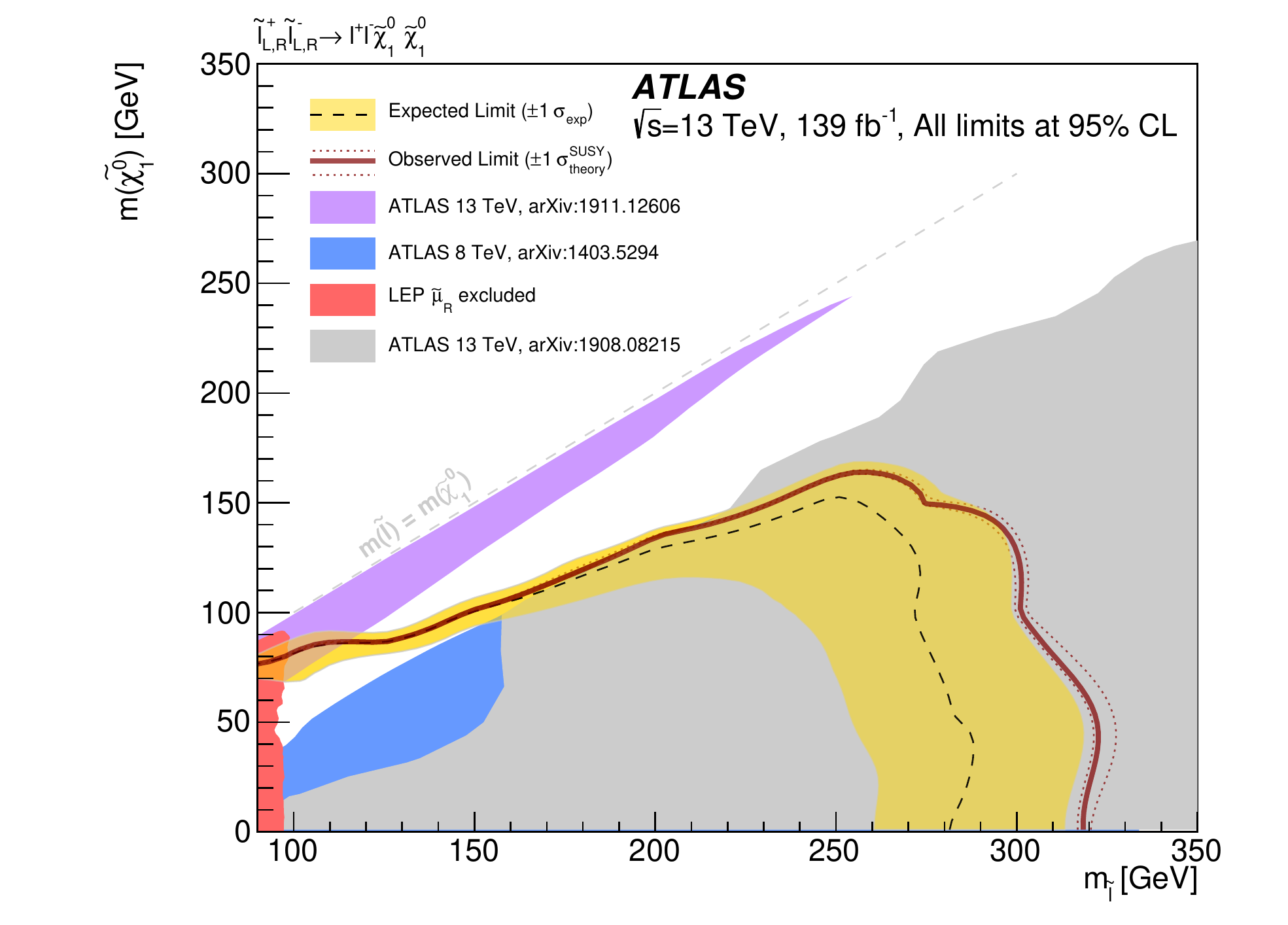}
\caption{}
\end{subfigure}
\begin{subfigure}[b]{\linewidth}
\centering
\includegraphics[width=0.7\linewidth]{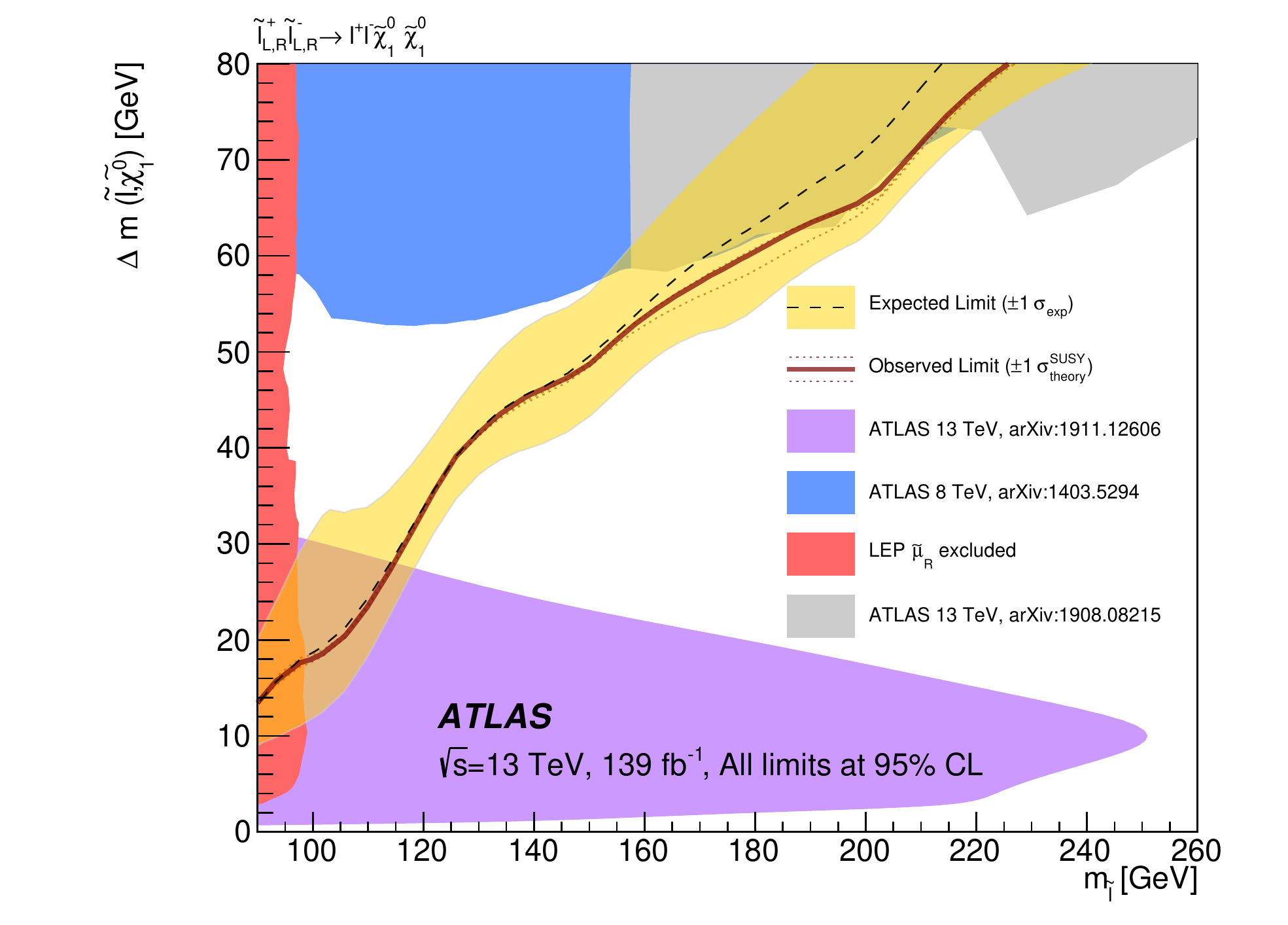}
\caption{}
\end{subfigure}
\caption{Observed and expected exclusion limits on SUSY simplified models for slepton-pair production in the (a) $m(\tilde{\ell})$--$m(\tilde{\chi}_1^0)$ and (b) $m(\tilde{\ell})$--$\Delta m(\tilde{\ell},\tilde{\chi}_1^0)$ planes. Only $\tilde{e}$ and $\tilde{\mu}$ are considered. The observed (solid thick line) and expected (thin dashed line) exclusion contours are shown. The shaded band around the dashed line corresponds to the $\pm 1 \sigma $ variations of the expected limit, including all uncertainties except theoretical uncertainties in the signal cross-section. The dotted lines around the observed limit illustrate the change in the observed limit as the nominal signal cross-section is scaled up and down by the theoretical uncertainty. All limits are computed at 95\% CL\@. The observed limits obtained at LEP~\cite{LEP-slepton} for $\tilde{\mu}_{\textup{R}}$ and by the ATLAS experiment in previous searches are also shown~\cite{SUSY-2018-16,SUSY-2013-11,SUSY-2018-32}.
}
\label{fig:exclusion_slep}
\end{figure}
 
\end{DIFnomarkup}

\begin{DIFnomarkup}
\begin{figure}[htb!]
\centering
\begin{subfigure}[]{0.48\linewidth}
\centering
\includegraphics[width=\linewidth]{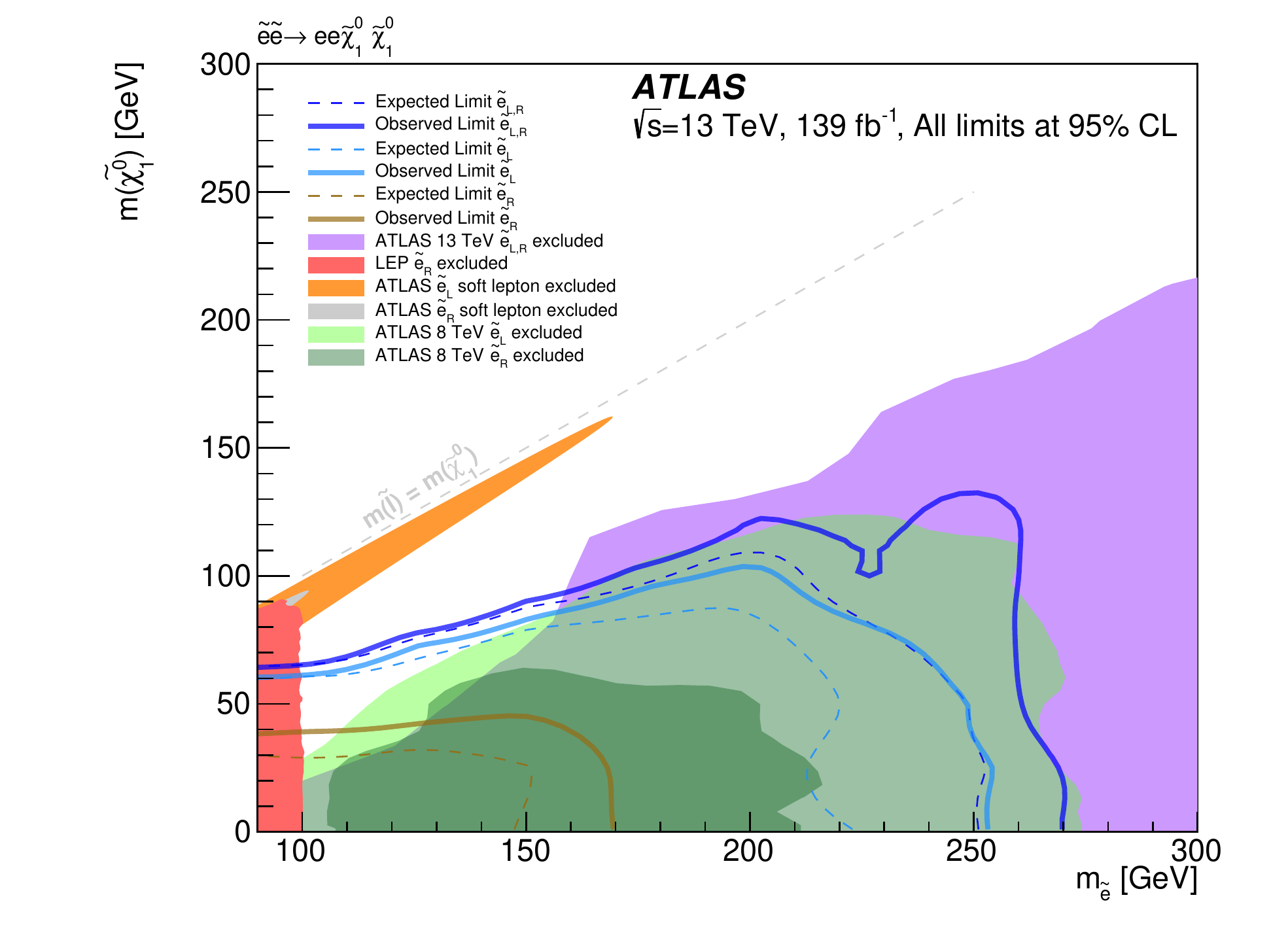}
\caption{}
\end{subfigure}
\begin{subfigure}[]{0.48\linewidth}
\centering
\includegraphics[width=\linewidth]{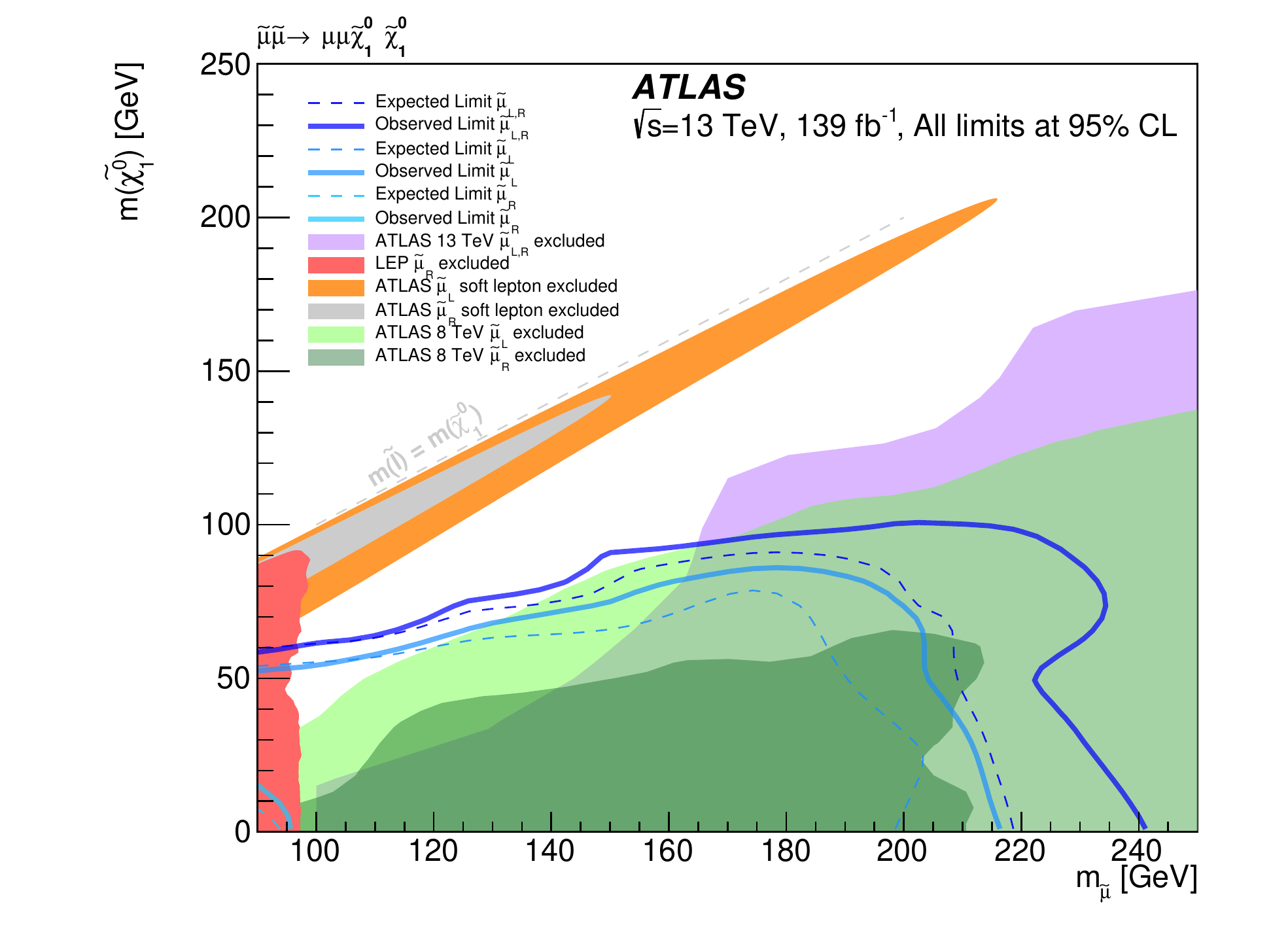}
\caption{}
\end{subfigure}
\begin{subfigure}[]{0.48\linewidth}
\centering
\includegraphics[width=\linewidth]{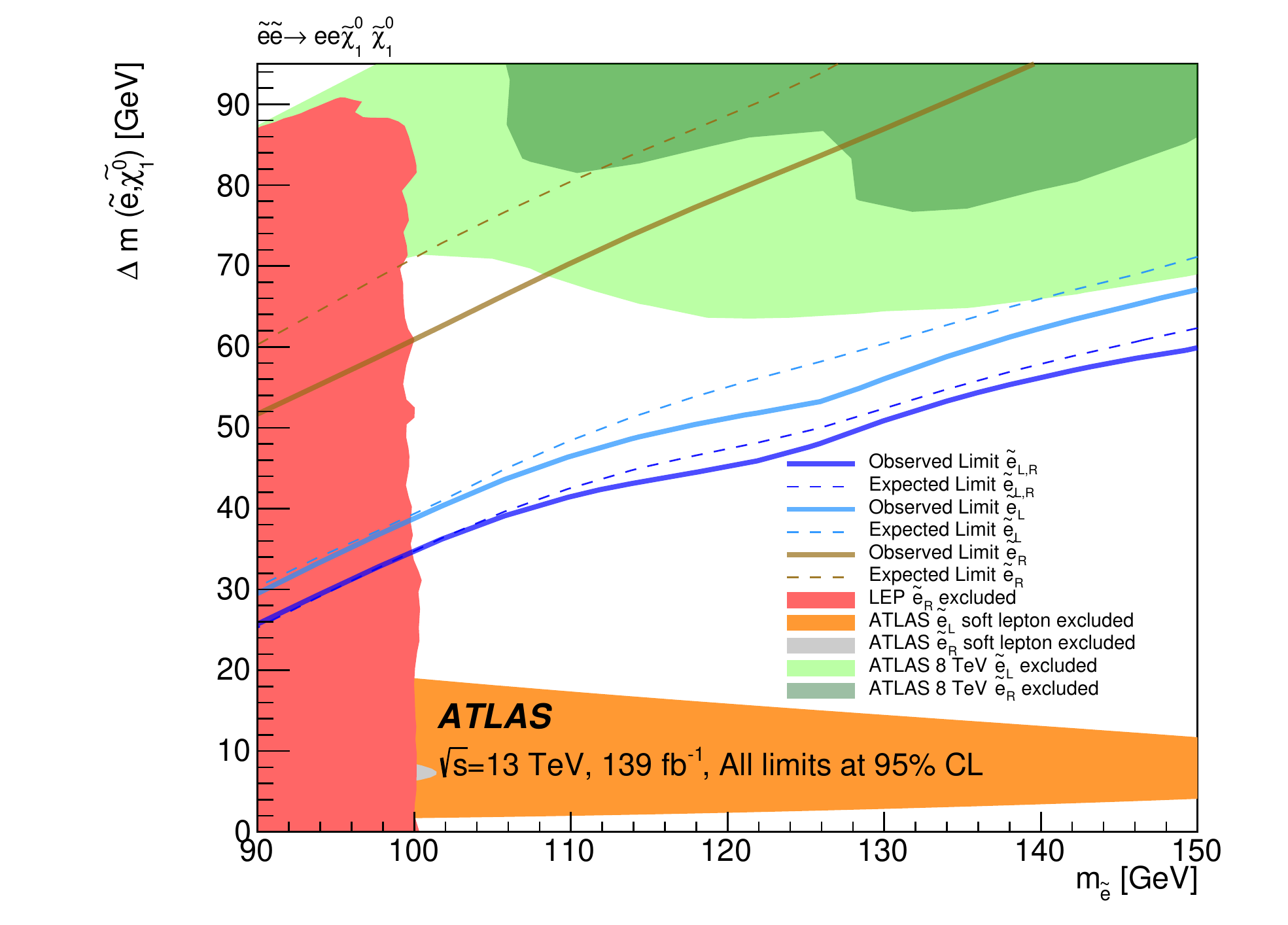}
\caption{}
\end{subfigure}
\begin{subfigure}[]{0.48\linewidth}
\centering
\includegraphics[width=\linewidth]{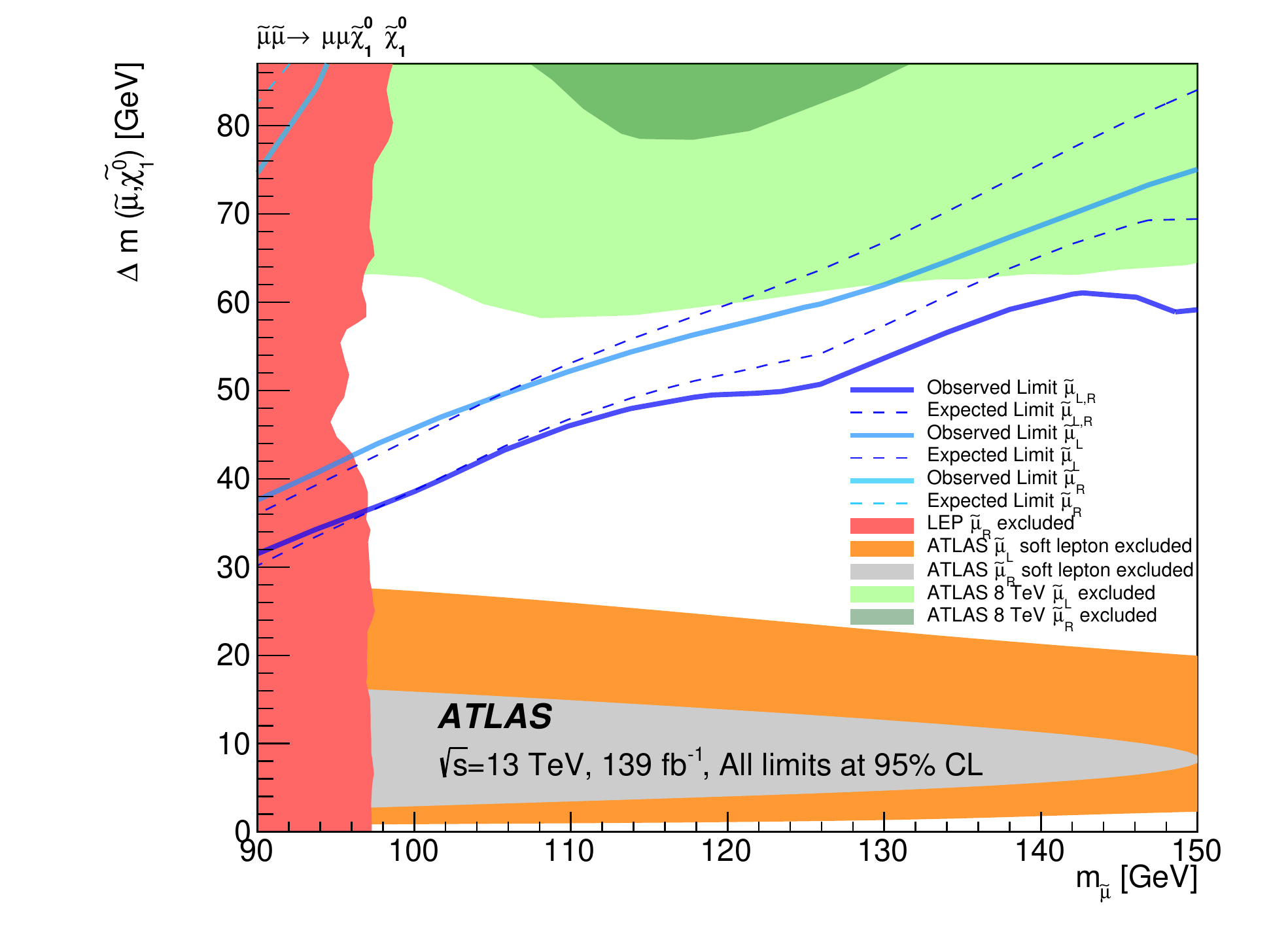}
\caption{}
\end{subfigure}
 
\caption{
Observed and expected exclusion limits on SUSY simplified models for direct selectron production in the (a) $m(\tilde{e})$--$m(\tilde{\chi}_1^0)$ and (c) $m(\tilde{e})$--$\Delta m(\tilde{e},\tilde{\chi}_1^0)$ planes, and for direct smuon production in the (b) $m(\tilde{\mu})$--$m(\tilde{\chi}_1^0)$ and (d) $m(\tilde{\mu})$--$\Delta m(\tilde{\mu},\tilde{\chi}_1^0)$ planes.
In figures (a) and (c) the observed (solid thick lines) and expected (dashed lines) exclusion contours are shown for combined $\tilde{e}_{\textup{L,R}}$ and for $\tilde{e}_{\textup{L}}$ and $\tilde{e}_{\textup{R}}$.
In figures (b) and (d) the observed (solid thick lines) and expected (dashed lines) exclusion contours are shown for combined $\tilde{\mu}_{\textup{L,R}}$ and for $\tilde{\mu}_{\textup{L}}$. No unique sensitivity to $\tilde{\mu}_{\textup{R}}$ is observed.
All limits are computed at 95\% CL.\@
The observed limits obtained at LEP~\cite{LEP-slepton} and by the ATLAS experiment in previous searches are also shown in the shaded areas~\cite{SUSY-2018-16,SUSY-2013-11,SUSY-2018-32}.
}
 
\label{fig:sleptonExclusions}
\end{figure}
\end{DIFnomarkup}

\FloatBarrier
 
\subsection{Results of the chargino search}

\begin{figure}[h]
\centering
\includegraphics[width=\linewidth]{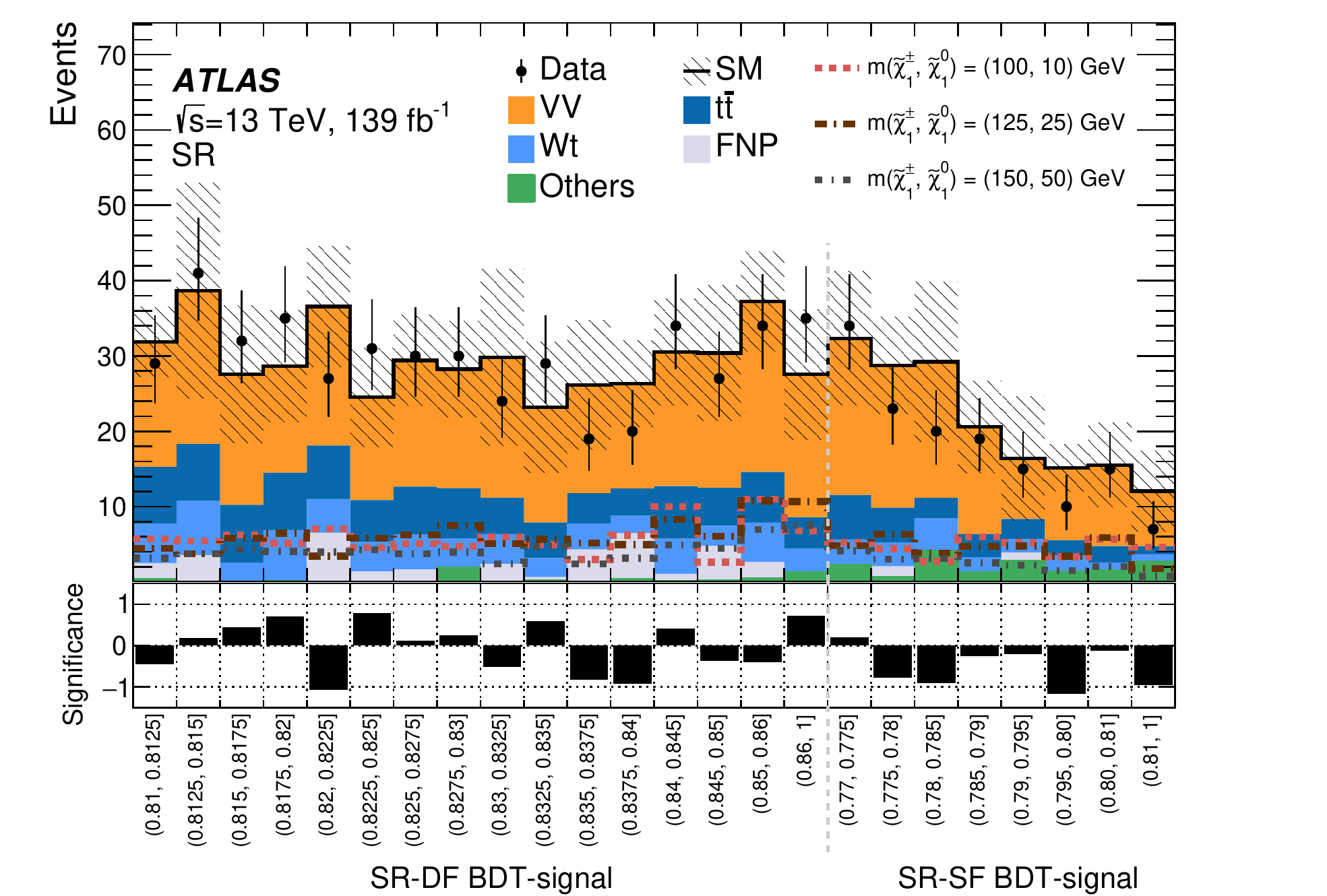}
\caption{The upper panel shows the observed number of events in the SRs defined in Table~\ref{tab:SRchargdefs}, together with the expected SM backgrounds obtained after the background fit in the CRs. `Others' include the non-dominant background sources, e.g.\ $t \bar{t}$+$V$, Higgs boson and Drell--Yan events. The uncertainty band includes systematic and statistical errors from all sources. Distributions for three benchmark signal points are overlaid for comparison. The lower panel shows the significance as defined in Ref.~\cite{Cousin:2008clt}.
}
\label{fig:SRpull_charg}
\end{figure}

The predicted numbers of background events, obtained from the results of the background fit in the binned SRs defined in Table~\ref{tab:SRchargdefs}, are shown together with the observed data in Figure~\ref{fig:SRpull_charg}.
The observed and predicted numbers of background events in the inclusive SRs are shown in Table~\ref{tab:SRinc_charg}.
The model-independent upper limits on the visible signal cross-section $\sigma^\mathrm{obs}$,
the observed and expected limits at 95\% CL on the number of potential beyond-the-SM events, and the $p_0$-values for each inclusive SR are also reported.
No significant deviations from the SM expectations are observed in any of the SRs considered, as shown in Figure~\ref{fig:SRpull_charg}.
 
\begin{figure}[htb!]
\centering
\begin{subfigure}[a]{\linewidth}
\centering
\includegraphics[width=0.7\linewidth]{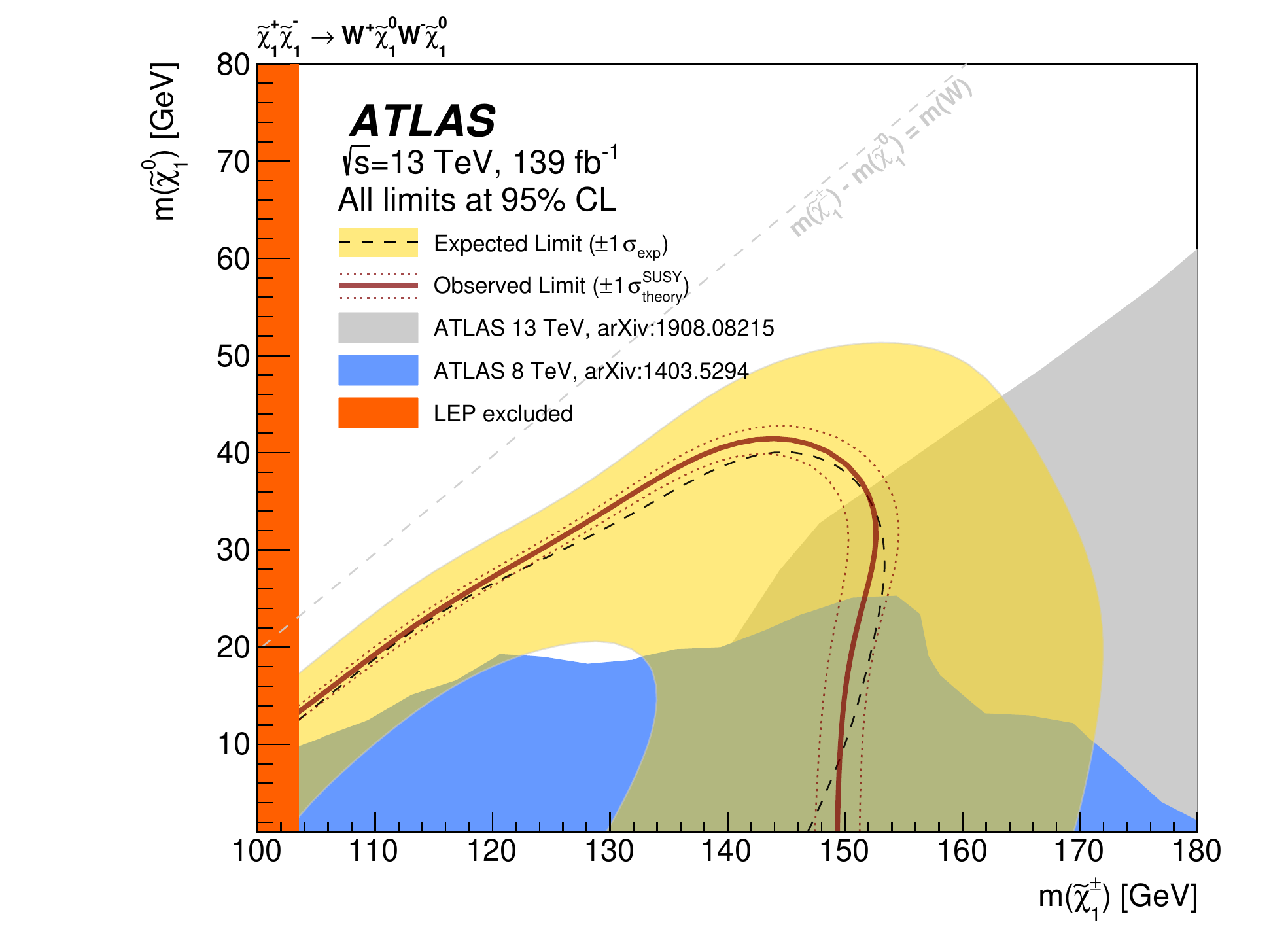}
\caption{}
\end{subfigure}
\begin{subfigure}[b]{\linewidth}
\centering
\includegraphics[width=0.7\linewidth]{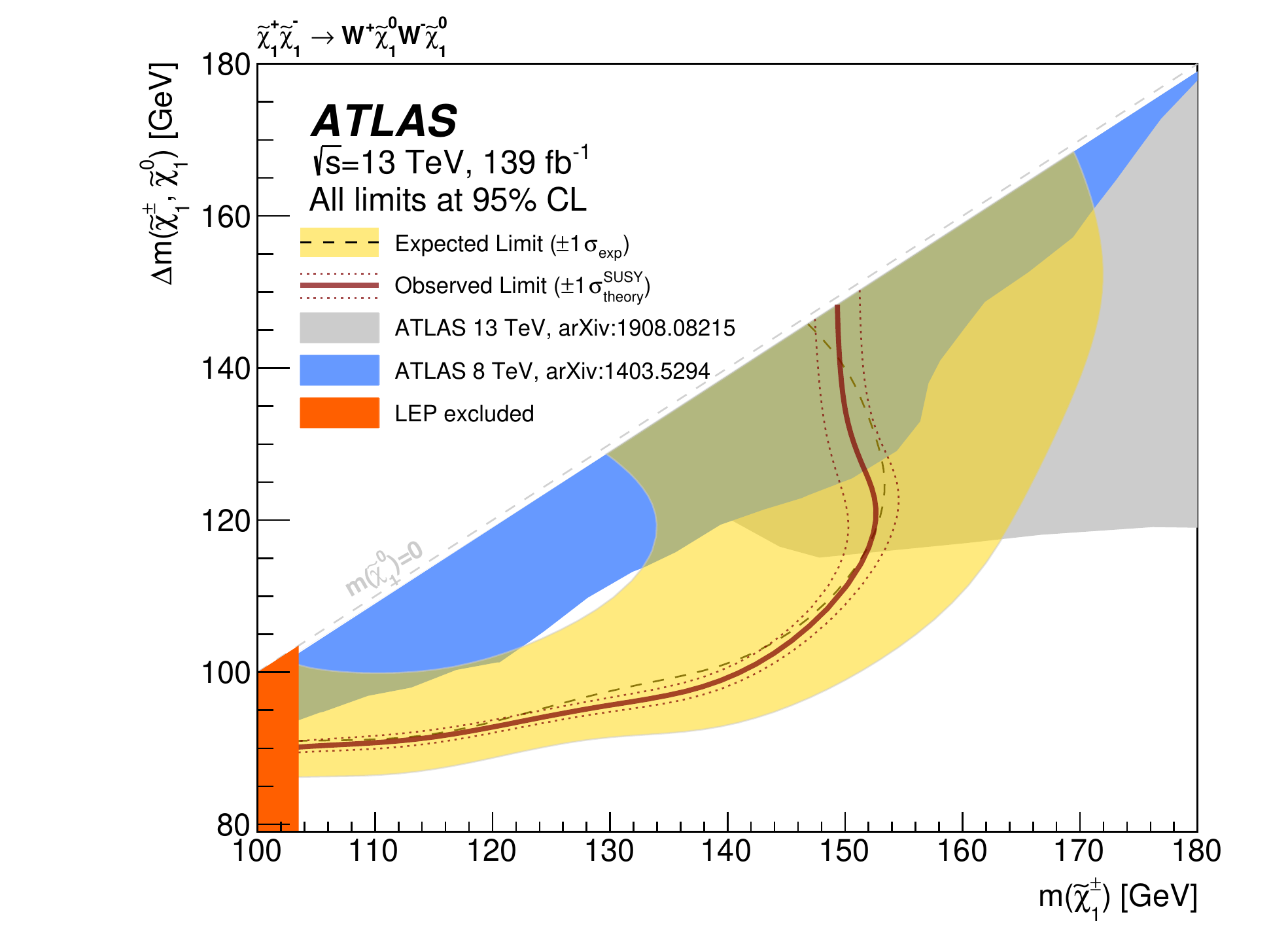}
\caption{}
\end{subfigure}
\caption{Observed and expected exclusion limits on SUSY simplified models for chargino-pair production with $W$-boson-mediated decays in the (a) $m(\tilde{\chi}_1^\pm)$--$m(\tilde{\chi}_1^0)$ and (b) $m(\tilde{\chi}_1^\pm)$--$\Delta m(\tilde{\chi}_1^\pm,\tilde{\chi}_1^0)$ planes. The observed (solid thick line) and expected (thin dashed line) exclusion contours are shown. The shaded band around the dashed line corresponds to the $\pm 1 \sigma $ variations of the expected limit, including all uncertainties except theoretical uncertainties in the signal cross-section. The dotted lines around the observed limit illustrate the change in the observed limit as the nominal signal cross-section is scaled up and down by the theoretical uncertainty. All limits are computed at 95\% CL\@. The observed limits obtained at LEP~\cite{LEP-chargino} and by the ATLAS experiment in previous 
searches are also shown~\cite{SUSY-2018-32,SUSY-2013-11}. In the case of the search performed on ATLAS Run~1 data at $\sqrt{s} = 8$~\TeV~\cite{SUSY-2013-11} no sensitivity was expected for the exclusion in the mass plane.
}
\label{fig:exclusion_charg}
\end{figure}

Exclusion limits at 95\% CL are set, using the CL$_{\text{s}}$ prescription, on the masses of the chargino and the LSP. These
include the exclusive SRs and the CRs in the simultaneous likelihood fit. The SF and DF SRs are included in the likelihood fit. The exclusion limits are shown in Figure~\ref{fig:exclusion_charg}. Chargino masses up to 140~\GeV\ are excluded at 95\% CL in the case of a mass-splitting between the chargino and neutralino as low as about 100~\GeV.

\begin{table}
\caption{Observed event yields and predicted background yields for the inclusive SRs defined in Table~\ref{tab:SRchargdefs} are reported in the left part of the table, as obtained using the background fit described in Section~\ref{sec:result}. The right part shows the model-independent upper limits at 95\% CL on the observed and expected numbers of beyond-the-SM events $S^{0.95}_\mathrm{obs/exp}$ and on the effective beyond-the-SM cross-section $\sigma^\mathrm{obs}$ ($\langle\mathrm{A}\epsilon\sigma\rangle^{0.95}_{\mathrm{obs}}$). They were obtained using a fit which includes the SRs, also described in Section~\ref{sec:result}. The $\pm 1 \sigma$ variations of $S^{0.95}_\mathrm{exp}$ are provided. The last column shows the $p_0$-value of the SM-only hypothesis.
For SRs where the data yield is smaller than expected, the $p_0$-value is capped at 0.50.}
\label{tab:SRinc_charg}
\begin{center}{
\setlength{\tabcolsep}{0.0pc}
\bgroup
\def\arraystretch{1.5}
\begin{tabular*}{\textwidth}{@{\extracolsep{\fill}}lcc@{\hskip 0.3in} | cccc} 
\noalign{\smallskip}\hline\noalign{\smallskip}{\textbf{ Signal region}}        & Observed & Expected     & $\sigma^\mathrm{obs}$ [fb]  &  $S_{\textrm{obs}}^{0.95}$  & $S_{\textrm{exp}}^{0.95}$ &  $p_0$  \\
\noalign{\smallskip}\hline\noalign{\smallskip}
SR$^{\text{-DF BDT-signal}\in(0.81,1]}_{\text{-SF BDT-signal}\in(0.77,1]}$    & $620$  &$630 \pm 70$  &$1.20$ &  $166.2$~~ & $ { 175.1 }^{ +44.9 }_{ -49.2 }$~~ & $ 0.50$ \\%
SR-DF BDT-signal$\in$(0.81,1]   &$477$  &$470 \pm 50$  & $0.80$ &  $111.0$~~ & $ { 108.9 }^{ +43.1 }_{ -31.1 }$~~ & $ 0.47$\\%
SR-DF BDT-signal$\in$(0.82,1]   &$340$  & $350 \pm 40$ & $0.55$ &  $76.0$ & $ { 81.5 }^{ +32.7 }_{ -22.9 }$& $ 0.50$\\%
SR-DF BDT-signal$\in$(0.83,1]   & $222$ & $231 \pm 26$ &  $0.38$ &  $52.3$ & $ { 57.8 }^{ +22.9 }_{ -16.1 }$ & $ 0.50$\\%
SR-DF BDT-signal$\in$(0.84,1]   & $130$ &  $126 \pm 15$& $0.29$ &  $40.0$ & $ { 37.5 }^{ +15.0 }_{ -10.5 }$ & $ 0.41$\\%
SR-DF BDT-signal$\in$(0.85,1]   & ~~$69$ & ~~$65 \pm 10$ & $0.22$ &  $30.9$ & $ { 28.0 }^{ +12.0 }_{ -8.3 }$ & $ 0.38$\\%
SR-SF BDT-signal$\in$(0.77,1]   &$143$  &$167 \pm 32$  & $0.47$ &  $65.5$ & $ { 80.6 }^{ +19.4 }_{ -23.0 }$  & $ 0.50$ \\%
SR-SF BDT-signal$\in$(0.78,1]   & ~~$86$ & $108 \pm 23$  & $0.31$ &  $42.8$ & $ { 53.9 }^{ +18.9 }_{ -13.6 }$  & $ 0.50$\\%
SR-SF BDT-signal$\in$(0.79,1]   & ~~$47$ & ~~$58 \pm 15$ & $0.21$ &  $28.9$ & $ { 34.1 }^{ +10.8}_{ -7.8 }$  & $ 0.50$\\%
SR-SF BDT-signal$\in$(0.80,1]   & ~~$22$  & $28 \pm 8$  &  $0.10$ &  $14.3$ & $ { 16.8 }^{ +5.9 }_{ -4.5 }$~ & $ 0.50$\\%
\noalign{\smallskip}\hline\noalign{\smallskip}
\end{tabular*}
\egroup
}
\end{center}
\end{table}

\FloatBarrier

\section{Conclusion}
\label{sec:conclusion}
 
The results of a search for the electroweak production of charginos and sleptons decaying into final states containing two leptons with opposite electric charge and missing transverse momentum are presented. The search uses 139~fb$^{-1}$ of $\sqrt{s}=13$~\TeV\ proton--proton collisions collected by the ATLAS experiment at the LHC during Run~2 (2015--2018).
Two scenarios are considered: the direct production of slepton pairs, where each slepton decays directly into the lightest neutralino and a lepton, and the production of lightest-chargino pairs, where each decays into a final state with the lightest neutralino plus a lepton via a $W$-boson decay.
The regions with mass differences up to approximately 150~\GeV\ between the sleptons and neutralino  and between the chargino and neutralino are explored in these analyses. Models with smuon production with mass differences in this region of the $m(\tilde{\mu})$--$m(\tilde{\chi}_1^0)$ plane are favoured to explain the $(g-2)_\mu$ anomaly for small $\tan\beta$ values. Their decay topologies are similar to those of SM processes, making it challenging to separate signal from background.  In order to target these models, a data-driven technique is used to estimate the main backgrounds in the slepton search, and a semi-data-driven approach using CRs to normalize the main backgrounds, classified with a BDT, is used in the chargino search.
 
The data are found to be consistent with the Standard Model predictions and exclusion limits at 95\% CL are set on the masses of relevant supersymmetric particles in each of these scenarios. Slepton masses up to 150~\GeV\ are excluded at 95\% CL in the case of a 50~\GeV\ mass-splitting between the sleptons and neutralino, thus surpassing the exclusion limits previously set by the LEP experiments~\cite{LEP-slepton}, and chargino masses up to 140~\GeV\ are excluded at 95\% CL in the case of a mass-splitting between the chargino and neutralino as low as about 100~\GeV. Compared to previous analyses for the same scenarios,
in the regions with a mass difference up to about 150~\GeV\ between the slepton or chargino and neutralino,
the results of these searches extend beyond the exclusion limits previously set at the LHC by ATLAS and CMS~\cite{SUSY-2013-11,SUSY-2016-24,CMS-SUS-13-006,CMS-SUS-14-002,CMS-SUS-16-039,CMS-SUS-17-004,CMS-SUS-17-010}.

\section*{Acknowledgements}

% The next lines are included from the .//acknowledgements/Acknowledgements.tex input file

We thank CERN for the very successful operation of the LHC, as well as the
support staff from our institutions without whom ATLAS could not be
operated efficiently.
 
We acknowledge the support of
ANPCyT, Argentina;
YerPhI, Armenia;
ARC, Australia;
BMWFW and FWF, Austria;
ANAS, Azerbaijan;
CNPq and FAPESP, Brazil;
NSERC, NRC and CFI, Canada;
CERN;
ANID, Chile;
CAS, MOST and NSFC, China;
Minciencias, Colombia;
MEYS CR, Czech Republic;
DNRF and DNSRC, Denmark;
IN2P3-CNRS and CEA-DRF/IRFU, France;
SRNSFG, Georgia;
BMBF, HGF and MPG, Germany;
GSRI, Greece;
RGC and Hong Kong SAR, China;
ISF and Benoziyo Center, Israel;
INFN, Italy;
MEXT and JSPS, Japan;
CNRST, Morocco;
NWO, Netherlands;
RCN, Norway;
MEiN, Poland;
FCT, Portugal;
MNE/IFA, Romania;
MESTD, Serbia;
MSSR, Slovakia;
ARRS and MIZ\v{S}, Slovenia;
DSI/NRF, South Africa;
MICINN, Spain;
SRC and Wallenberg Foundation, Sweden;
SERI, SNSF and Cantons of Bern and Geneva, Switzerland;
MOST, Taiwan;
TENMAK, T\"urkiye;
STFC, United Kingdom;
DOE and NSF, United States of America.
In addition, individual groups and members have received support from
BCKDF, CANARIE, Compute Canada and CRC, Canada;
PRIMUS 21/SCI/017 and UNCE SCI/013, Czech Republic;
COST, ERC, ERDF, Horizon 2020 and Marie Sk{\l}odowska-Curie Actions, European Union;
Investissements d'Avenir Labex, Investissements d'Avenir Idex and ANR, France;
DFG and AvH Foundation, Germany;
Herakleitos, Thales and Aristeia programmes co-financed by EU-ESF and the Greek NSRF, Greece;
BSF-NSF and MINERVA, Israel;
Norwegian Financial Mechanism 2014-2021, Norway;
NCN and NAWA, Poland;
La Caixa Banking Foundation, CERCA Programme Generalitat de Catalunya and PROMETEO and GenT Programmes Generalitat Valenciana, Spain;
G\"{o}ran Gustafssons Stiftelse, Sweden;
The Royal Society and Leverhulme Trust, United Kingdom.
 
The crucial computing support from all WLCG partners is acknowledged gratefully, in particular from CERN, the ATLAS Tier-1 facilities at TRIUMF (Canada), NDGF (Denmark, Norway, Sweden), CC-IN2P3 (France), KIT/GridKA (Germany), INFN-CNAF (Italy), NL-T1 (Netherlands), PIC (Spain), ASGC (Taiwan), RAL (UK) and BNL (USA), the Tier-2 facilities worldwide and large non-WLCG resource providers. Major contributors of computing resources are listed in Ref.~\cite{ATL-SOFT-PUB-2021-003}.

% End of text imported from the .//acknowledgements/Acknowledgements.tex input file

\clearpage
 
\printbibliography
 
\clearpage
% ATLAS Collaboration author list
% Reference date of SUSY-2019-02 is 2022-02-17
% Author list last updated on date 11-JUL-23
% Data extracted on 11-Jul-2023 for paper reference SUSY-2019-02
% at 11:18am
 
\begin{flushleft}
\hypersetup{urlcolor=black}
{\Large The ATLAS Collaboration}

\bigskip

\AtlasOrcid[0000-0002-6665-4934]{G.~Aad}$^\textrm{\scriptsize 101}$,
\AtlasOrcid[0000-0002-5888-2734]{B.~Abbott}$^\textrm{\scriptsize 119}$,
\AtlasOrcid[0000-0002-7248-3203]{D.C.~Abbott}$^\textrm{\scriptsize 102}$,
\AtlasOrcid[0000-0002-1002-1652]{K.~Abeling}$^\textrm{\scriptsize 55}$,
\AtlasOrcid[0000-0002-8496-9294]{S.H.~Abidi}$^\textrm{\scriptsize 29}$,
\AtlasOrcid[0000-0002-9987-2292]{A.~Aboulhorma}$^\textrm{\scriptsize 35e}$,
\AtlasOrcid[0000-0001-5329-6640]{H.~Abramowicz}$^\textrm{\scriptsize 150}$,
\AtlasOrcid[0000-0002-1599-2896]{H.~Abreu}$^\textrm{\scriptsize 149}$,
\AtlasOrcid[0000-0003-0403-3697]{Y.~Abulaiti}$^\textrm{\scriptsize 116}$,
\AtlasOrcid[0000-0003-0762-7204]{A.C.~Abusleme~Hoffman}$^\textrm{\scriptsize 136a}$
% ...
\bigskip
\\

% ...
$^{29}$Physics Department, Brookhaven National Laboratory, Upton NY; United States of America.\\
% ...
$^{35}$$^{(a)}$Facult\'e des Sciences Ain Chock, R\'eseau Universitaire de Physique des Hautes Energies - Universit\'e Hassan II, Casablanca;$^{(b)}$Facult\'{e} des Sciences, Universit\'{e} Ibn-Tofail, K\'{e}nitra;$^{(c)}$Facult\'e des Sciences Semlalia, Universit\'e Cadi Ayyad, LPHEA-Marrakech;$^{(d)}$LPMR, Facult\'e des Sciences, Universit\'e Mohamed Premier, Oujda;$^{(e)}$Facult\'e des sciences, Universit\'e Mohammed V, Rabat;$^{(f)}$Institute of Applied Physics, Mohammed VI Polytechnic University, Ben Guerir; Morocco.\\
% ...
$^{55}$II. Physikalisches Institut, Georg-August-Universit\"{a}t G\"ottingen, G\"ottingen; Germany.\\
% ...
$^{101}$CPPM, Aix-Marseille Universit\'e, CNRS/IN2P3, Marseille; France.\\
$^{102}$Department of Physics, University of Massachusetts, Amherst MA; United States of America.\\
% ...
$^{116}$Department of Physics, New York University, New York NY; United States of America.\\
% ...
$^{119}$Homer L. Dodge Department of Physics and Astronomy, University of Oklahoma, Norman OK; United States of America.\\
% ...
$^{136}$$^{(a)}$Departamento de F\'isica, Pontificia Universidad Cat\'olica de Chile, Santiago;$^{(b)}$Millennium Institute for Subatomic physics at high energy frontier (SAPHIR), Santiago;$^{(c)}$Instituto de Investigaci\'on Multidisciplinario en Ciencia y Tecnolog\'ia, y Departamento de F\'isica, Universidad de La Serena;$^{(d)}$Universidad Andres Bello, Department of Physics, Santiago;$^{(e)}$Instituto de Alta Investigaci\'on, Universidad de Tarapac\'a, Arica;$^{(f)}$Departamento de F\'isica, Universidad T\'ecnica Federico Santa Mar\'ia, Valpara\'iso; Chile.\\
% ...
$^{149}$Department of Physics, Technion, Israel Institute of Technology, Haifa; Israel.\\
$^{150}$Raymond and Beverly Sackler School of Physics and Astronomy, Tel Aviv University, Tel Aviv; Israel.\\
% ...
$^{a}$ Also Affiliated with an institute covered by a cooperation agreement with CERN.\\
$^{b}$ Also at Borough of Manhattan Community College, City University of New York, New York NY; United States of America.\\
$^{c}$ Also at Bruno Kessler Foundation, Trento; Italy.\\
$^{d}$ Also at Center for High Energy Physics, Peking University; China.\\
$^{e}$ Also at Centro Studi e Ricerche Enrico Fermi; Italy.\\
$^{f}$ Also at CERN, Geneva; Switzerland.\\
% ...

\end{flushleft}

% Created with Glance <Atlas.Glance@cern.ch>

\clearpage

\end{document}